\pdfoutput=1
\documentclass[10pt,twocolumn,letterpaper]{article}

\usepackage{iccv}
\usepackage{times}
\usepackage{epsfig}
\usepackage{graphicx}
\usepackage{amsmath}
\usepackage{amssymb}
\usepackage{algorithm}
\usepackage[noend]{algpseudocode}
\usepackage[nice]{nicefrac}
\usepackage{multirow}

\usepackage[breaklinks=true,bookmarks=false]{hyperref}

\iccvfinalcopy % *** Uncomment this line for the final submission

\def\iccvPaperID{25} % *** Enter the ICCV Paper ID here

\def\NoNumber#1{{\def\alglinenumber##1{}\State#1}\addtocounter{ALG@line}{-1}}

\ificcvfinal\fi

\begin{document}

%%%%%%%%% TITLE
\title{Bi-Directional ConvLSTM U-Net with Densley Connected Convolutions}

\author{Reza Azad\textsuperscript{*}\\
Sharif University \\
of Tech., Iran\\
{\tt\small rezazad68@gmail.com}
\and
Maryam Asadi-Aghbolaghi\thanks{These two authors contributed equally.}\\
Inst. for Research in\\
Fundamental Sciences (IPM), Iran\\
{\tt\small masadi@ipm.ir}
\and
Mahmood Fathy\\
Iran University of\\
Science and Tech., Iran\\
{\tt\small mahfathy@iust.ac.ir}
\and
Sergio Escalera\\
Universitat de Barcelona and \\
Computer Vision Center, Spain\\
{\tt\small sergio@maia.ub.es}
}

\maketitle
\ificcvfinal\thispagestyle{empty}\fi

%%%%%%%%% ABSTRACT
\begin{abstract}
In recent years, deep learning-based networks have achieved state-of-the-art performance in medical image segmentation. Among the existing networks, U-Net has been successfully applied on medical image segmentation.
In this paper, we propose an extension of U-Net, Bi-directional ConvLSTM U-Net with Densely connected convolutions (BCDU-Net), for medical image segmentation, in which we take full advantages of U-Net, bi-directional ConvLSTM (BConvLSTM) and the mechanism of dense convolutions. Instead of a simple concatenation in the skip connection of U-Net, we employ BConvLSTM to combine the feature maps extracted from the corresponding encoding path and the previous decoding up-convolutional layer in a non-linear way. 
%To capture more global information, we also augment the last convolutional layer of the encoding path with a self-attention mechanism. 
To strengthen feature propagation and encourage feature reuse, we use densely connected convolutions in the last convolutional layer of the encoding path.  
Finally, we can accelerate the convergence speed of the proposed network by employing batch normalization (BN). The proposed model is evaluated on three datasets of: retinal blood vessel segmentation, skin lesion segmentation, and lung nodule segmentation, achieving state-of-the-art performance.
\end{abstract}

\section{Introduction}

Medical images play a key role in medical treatment and diagnosis. The goal of Computer-Aided Diagnosis (CAD) systems is providing doctors with precise interpretation of medical images to have better treatment of a large number of people. %The medical imaging includes different kinds of imaging techniques like Magnetic Resonance Imaging (MRI), ultrasound, and Computer Tomography (CT). 
Moreover, automatic processing of medical images results in reducing the time, cost, and error of human-based processing. One of the main research areas in this field is medical image segmentation, being a critical step in numerous medical imaging studies. 
Like other fields of research in computer vision, deep learning networks achieve outstanding results and use to outperform non-deep state-of-the-art methods in medical imaging. 
Deep neural networks are mostly utilized in classification tasks, where the output of the network is a single label or probability values associated labels to a given input image. %AlexNet, VGG, Residual Net, and DenseNet are among the networks for classification task. 
These networks work fine thanks to some structural features \cite{alom2018} such as: activation function, different efficient optimization algorithms, and dropout as a regularizer for the network. 
These networks require a large amount of data to train and provide a good generalization behavior given the huge number of network parameters. A critical issue in medical image segmentation is the unavailability of large (and annotated) datasets. In medical image segmentation, per pixel labeling is required instead of image level label. 

%%%%To mitigate this problem, many local regions of the input image are selected and utilized as the input data to networks.
Fully convolutional neural network (FCN) \cite{long2015fully} was one of the first deep networks applied to image segmentation. Ronneberger et al. \cite{ronneberger2015} extended this architecture to U-Net, achieving good segmentation results leveraging the need of a large amount of training data. Their network consists of  encoding and decoding paths. In the encoding path a large number of feature maps with reduced dimensionality are extracted from the input data. The decoding path is used to produce segmentation maps (with the same size as the input) by performing up-convolutions.
Many extensions of U-Net have been proposed so far \cite{alom2018,oktay2018}. The most important modification is mainly about the skipping connections. %In this part of the network, extracted feature maps from the previous up-convolutional layer are concatenated with the extracted features from the corresponding encoding path. 
In some extended versions of U-Net, the extracted feature maps in the skip connection are first fed to a processing step (e.g. attention gates \cite{oktay2018}) and then concatenated. The main drawback of these networks is that the processing step is performed individually for the two sets of feature maps, and these features are then simply concatenated. 

In this paper, we propose BCDU-Net, an extended version of the U-Net, by including BConvLSTM \cite{song2018pyramid} in the skip connection and reusing feature maps with densely convolutions. The feature maps from the corresponding encoding layer have higher resolution while the feature maps extracted from the previous up-convolutional layer contain more semantic information. Instead of a simple concatenation, combining these two kinds of feature maps with non-linear functions may result in more precise segmentation output. Therefore, in this paper we extend the U-Net architecture by adding BConvLSTM in the skip connection to combine these two kinds of feature maps.

Having a sequence of convolutional layers may help the network to learn more kinds of features; however, in many cases, the network learns redundant features. To mitigate this problem and enhance information flow through the network, we utilize the idea of densely connected convolutions \cite{huang2017densely}. In the last layer of the contracting path, convolutional blocks are connected to all subsequent blocks in that layer via channel-wise concatenation. Features which are learned in each block are passed forward to the next block. This strategy helps the method to learn a diverse set of features based on the “collective knowledge” gained by previous layers, and therefore, avoiding learning redundant features..

%Feature maps with the highest semantic information are produced in the last step of the encoding layer. It is well-known that convolutions just capture local information. In order to capture more global information, we augment the last convolutional operators in the encoding path with a self-attention mechanism, as proposed by Bello et al. \cite{bello2019attention}. To do that, the convolutional feature maps are concatenated with a set of feature maps produced via self-attention. 

Furthermore, we accelerate the convergence speed of the network by employing BN after the up-convolution filters. We evaluate the proposed BCDU-Net on three different applications of: retinal blood vessel segmentation (DRIVE datase), Skin lesion segmentation (ISIC 2018 dataset) and lung nodule segmentation (Lung dataset). The experimental results demonstrate that the proposed network achieves superior performance than state-of-the-art alternatives. \footnote{Source code is available on https://github.com/rezazad68/BCDU-Net.}

\section{Related Work}

One of the most crucial tasks in medical imaging is semantic segmentation. Before the revolution of deep learning in computer vision, traditional handcrafted features were exploited for semantic segmentation. During the last few years, deep learning-based approaches have outstandingly improved the performance of classical image segmentation strategies. % and recognition algorithms in different tasks of computer vision. 
%It is not very difficult to transfer these advances to medical images segmentation. 
Based on the exploited deep architecture, these approaches can be divided into three groups of: convolutional neural network (CNN), fully convolutional network (FCN), and recurrent neural network (RNN).

\subsection{Convolutional Neural Network (CNN)}
Cui et al. \cite{cui2016} exploited CNN for automatic segmentation of brain MRI images. The authors first divided the input images into some patches and then utilized these patches for training CNN. To handle an arbitrary number of modalities as the input data, Kleesiek et al. \cite{kleesiek2016} proposed a 3D CNN for brain lesion segmentation. To process MRI data, the network consists of four channels: non-enhanced and contrast-enhanced T1w, T2w and FLAIR contrasts. Roth et al. \cite{roth2015deeporgan} proposed a multi-level deep convolutional networks for pancreas segmentation in abdominal CT scans as a probabilistic bottom-up approach.

\subsection{Fully Convolutional Network (FCN)}
One of the main problems of the CNN models for segmentation tasks is that the spatial information of the image is lost when the convolutional features are fed into the fc layers. To overcome this problem the \emph{fully convolutional network} (FCN) was proposed by Long et al. \cite{long2015fully}. This network is trained end-to-end and pixels-to-pixels for semantic segmentation. All fc layers of the CNN architecture are replaced with convolutional and deconvolutional ones to keep the original spatial resolutions. Therefore, the original spatial dimension of the features maps are recovered while the network is performing the segmentation task.
FCN has been frequently utilized for segmentation of medical and biomedical images \cite{zhou2016three,zhou2017fixed}. 
Zhou et al. \cite{zhou2016three} exploited FCN for segmentation of anatomical structures on 3D CT images. An FCN with convolution and de-convolution parts is trained end-to-end, performing voxel-wise multiple-class classification to map each voxel in a CT image to an anatomical label. Drozdzal et al. \cite{drozdzal2016} proposed very deep FCN by using short skip connections. The authors showed that a very deep FCN with both long and short skip connections achieved better result than the original one.
%In that method the segmentation of the anatomical structures (including multiple organs) in a CT image (generally in 3D) is performed by majority voting of the semantic segmentation of multiple 2D slices drawn from different viewpoints.
%Roth et al. \cite{roth2018application} proposed to employ 3D FCN in a cascaded fashion for segmentation of the organs and vessels in CT images. The authors utilized a two-stage, coarse-to-fine architecture of FCN. In the first stage, about $40\%$ of the voxels (automatically generated mask of the patient’s body) are processed by FCN. This amount of voxels is reduced to $10\%$, and the network is allowed on more detailed segmentation.% In original FCN, long skip connections are utilized to skip features from the contracting path to the expanding one.

U-Net, proposed by Ronneberger et al. \cite{ronneberger2015}, is one of the most popular FCNs for medical image segmentation. This network consists of contracting and expanding paths. 
%In the contracting path, convolution filters followed by ReLU and $2\times2$ max-pooling operators are applied on the input data. 
%In the contracting path, convolution, ReLU and max-pooling operators are applied on the input data. In the expanding path, de-convolution are utilized to up-sample the feature maps. The feature maps were copied and cropped from encoding units to the decoding ones. 
U-Net has some advantages than the other segmentation-based network \cite{alom2018}. It works well with few training samples and the network is able to utilize the global location and context information at the same time.
%Different extension versions of U-Net have been proposed for segmentation task. 
Milletari et al. \cite{milletari2016} proposed V-Net, a 3D extension version of U-Net to predict segmentation of a given volume at once. In that network, the authors proposed an end-to-end 3D image segmentation network based on a volumetric (MRI volumes), fully convolutional, neural network. 
3D U-Net \cite{cciccek20163d} is proposed for processing 3D volumes instead of 2D images as input. In which, all 2D operations of U-Net are replaced with their 3D counterparts. 
%In that network, all 2D operations in U-Net are replaced with their 3D counterparts.%, i.e., 3D convolutions, 3D max pooling, and 3D up-convolutional layers. 
%In \cite{kayalibay2017}, two modifications have been applied on the standard U-Net. The authors combine multiple segmentation maps that are created at different scales. Moreover, to forward feature maps from one stage of the network to the other one, element-wise summation is utilized.
VoxResNet \cite{chen2016voxresnet}, a deep voxel-wise residual network, was proposed for brain segmentation from MR. This 3D network is inspired by deep residual learning, performing summation of feature maps from different layers. 
%A dual pathway 3D CNN (with 11 layers) \cite{kamnitsas2017} was proposed for brain lesion segmentation in multi-modal brain MRI. In this model, input images at multiple scales are fed simultaneously to a FCN. %Moreover, the adjacent image patches are processed into one pass through the network. The authors also employed a 3D fully connected Conditional Random Field which effectively removes false positives as a post-processing step.
%Li et al. proposed High-Res3DNet \cite{li2017compactness}, which is a high-resolution, compact convolutional network for volumetric image segmentation which include high spatial resolution feature maps throughout the layers.

\subsection{Recurrent Neural Network (RNN)}
%One of the most used neural networks for processing a sequence is RNN, which can take into account the temporal data using recurrent connections in hidden layers. It has been successfully applied for modeling short- and long-temporal sequences. These networks are able to model the global contexts and improve semantic segmentation.
%Different RNN based deep network have been proposed for semantic segmentation. 
Pinheiro et al. \cite{pinheiro2014} proposed an end-to-end feed forward deep network consisting of an RNN that can take into account long range label dependencies in the scenes while limiting the capacity of the model. Visin et al. \cite{visin2016reseg} proposed ReSeg for semantic segmentation. In that network, the input images are processed with a pre-trained VGG-16 model and its resulting feature maps are then fed into one or more ReNet layers. % Finally an up-sampling layer is employed for resizing the feature map.
DeepLab architecture \cite{chen2017deeplab} contains a deep convolutional neural network %which %is pre-trained in the task of image classification. 
in which all fully connected layers are replaced by convolutional layers and then the feature resolution is increased through atrous convolutional layers.
Alom et al. \cite{alom2018} proposed Recurrent Convolutional Neural Network (RCNN) and Recurrent Residual Convolutional Neural Network (R2CNN) based on U-Net models for medical image segmentation. %The residual unit helps the network in training. The feature accumulation with recurrent residual convolutional layers improve the feature representation for segmentation tasks.
Bai et al. \cite{bai2018recurrent} combined an FCN with an RNN for medical image sequence segmentation, which is able to incorporate both spatial and temporal information for MR images. %Gao \cite{gao2018fully} proposed an end to end combination of FCN and RNN with long short-term memory (LSTM) units for 4D segmentation of MRI images.

In this paper, BCDU-Net is proposed as an extension of U-Net, showing better performance than state-of-the-art alternatives for the segmentation task. Moreover, BN %utilized in the network 
has a significant effect on the convergence speed of the network.

\section{Proposed Method}
\begin{figure*}
\centering
%\vspace{-5mm}
\includegraphics[width=0.8\textwidth]{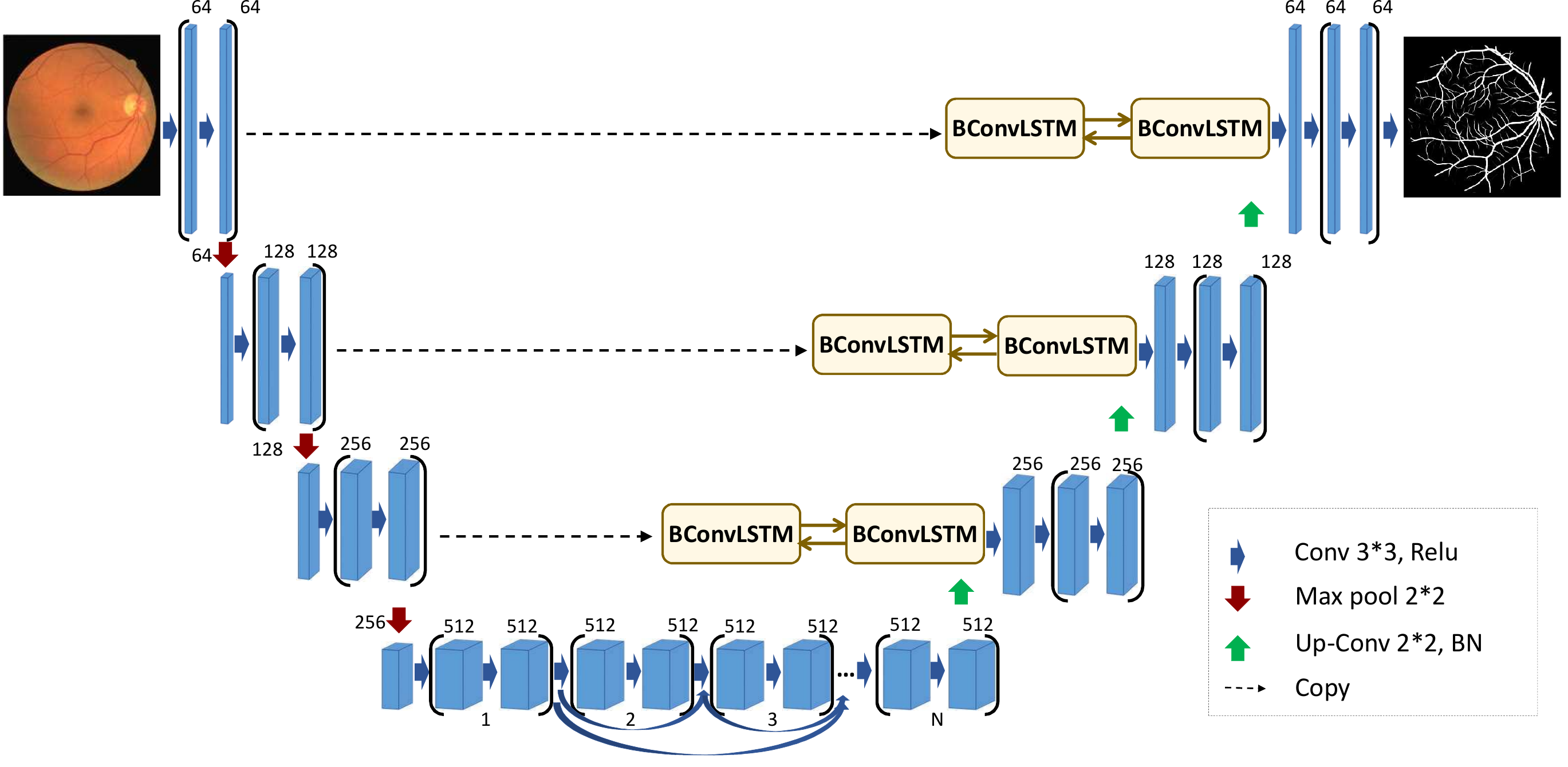}
%\vspace{-5mm}
\caption{BCDU-Net with bi-directional ConvLSTM in the skip connections and densely connected convolution.% in the last layer of contracting path.
} \label{fig:Model}
\vspace*{-\baselineskip}
\end{figure*}

Inspired by U-Net \cite{ronneberger2015}, BConvLSTM \cite{song2018pyramid}, and dense convolutions \cite{huang2017densely}, we propose the BCDU-Net as shown in Figure \ref{fig:Model}. The network utilizes the strengths of both BConvLSTM states and densely connected convolutions. %We highlight different parts of the proposed network with more details in the follwoing sub sections.
We detail different parts of the network in the next sub sections.
%It has been proved that augmenting a convolutional layer with a self attention mechanism \cite{bello2019attention} improves the classification rate. In this paper, we add a self attention-based convolution after the last convolutional layer in the encoding path. Moreover, in skip connection, instead of a simple concatenation of the corresponding cropped feature map from the contracting path and the result of up-convolutional layer, we employed ConvLSTM to process this data with more complex functions. Moreover, the outputs of the ConvLSTM are first normalized with a batch normalization layer (BN) and then fed to the next convolutional layer. We highlight different parts of the proposed network with more details into two sub-sections of encoding and decoding paths.

%\vspace{-7mm}

\subsection{Encoding Path}
The contracting path of BCDU-Net includes four steps. Each step consists of two convolutional $3\times3$ filters followed by a $2\times2$ max pooling function and ReLU. The number of feature maps are doubled at each step. %In fact, the convolutional layers in the contracting path process the local information layer by layers. 
The contracting path extracts progressively image representations and increases the dimension of these representations layer by layer. Ultimately, the final layer in the encoding path produces a high dimensional image representation with high semantic information.
The original U-Net contains a sequence of convolutional layers in the last step of encoding path.
Having a sequence of convolutional layers in a network yields the method learn different kinds of features. Nevertheless, the network might learn redundant features in the successive convolutions. To mitigate this problem, densely connected convolutions are proposed \cite{huang2017densely}. This helps the network to improve its performance by the idea of ``collective knowledge" in which the feature maps are reused through the network. It means feature maps learned from all previous convolutional layers are concatenated with the feature map learned from the current layer and then are forwarded to use as the input to the next convolution.

\begin{figure*}
\centering
%\vspace{-5mm}
\includegraphics[width=0.8\textwidth]{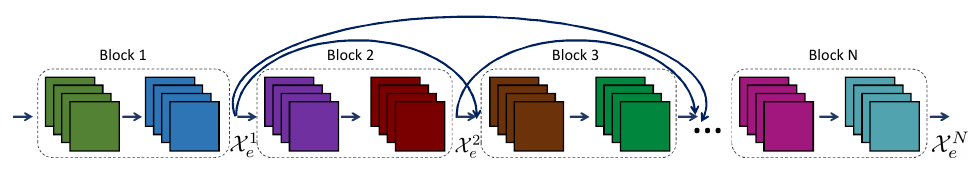}
%\vspace{-5mm}
\caption{Dense layer of the BCDU-Net.} \label{fig:dense}
\vspace*{-\baselineskip}
\end{figure*}

The idea of densely connected convolutions has some advantages over the regular convolutions \cite{huang2017densely}. First of all, it helps the network to learn a diverse set of feature maps instead of redundant features. Moreover, this idea improves the network's representational power by allowing information flow through the network and reusing features. Furthermore, dense connected convolutions can benefit from all the produced features before it, which prompt the network to avoid the risk of exploding or vanishing gradients. In addition, the gradients are sent to their respective places in the network more quickly in the backward path. 
We employ the idea of densely connected convolutions in the proposed network. To do that, we introduce one block as two consecutive convolutions. There are a sequence of $N$ blocks in the last convolutional layer of the encoding path, shown in Figure \ref{fig:dense}. These blocks are densely connected. % to each others. 
We consider $\mathcal{X}_e^i $ as the output of the $i^{th}$ convolutional block. The input of the $i^{th}$ ($i\in \{1,...,N\}$) convolutional block receives the concatenation of the feature maps of all preceding convolutional blocks as its input, i.e., $ \big[\mathcal{X}_e^{1},\mathcal{X}_e^{2}, ...,\mathcal{X}_e^{i-1} \big] \in \mathbb{R}^{(i-1)F_l\times W_l\times H_l}$ , and the output of the $i^{th}$ block is $\mathcal{X}_e^{i} \in \mathbb{R}^{F_l\times W_l\times H_l}$. In the remaining part of the paper we use simply $\mathcal{X}_e$ instead of $\mathcal{X}_e^{N}$.

%After the last convolutional layer, a self-attention based convolutional layer is included. The purpose of self-attention is to dynamically produce a weighted average of values computed from hidden units via a similarity function between them. Bello et al. \cite{bello2019attention} proposed to augment a convolutional layer with a self-attention, which maintains translation equivariance while being infused with relative position information. To do that, the convolutional feature maps are concatenated by the self-attention feature maps. %A number of attention maps (multihead attentions) are computed from a set of queries and keys for each spatial location of the input data. The output of these attention maps are a number of weighted averages of the values which are then concatenated and reshaped to match the original spatial size. The convolution is performed in parallel to multihead attention, and its feature maps are concatenated with the output of multihead attention. More information about the convolutional attention can be found in \cite{bello2019attention}.

\subsection{Decoding Path}
\begin{figure}
\centering
%\vspace{-5mm}
\includegraphics[width=0.5 \textwidth]{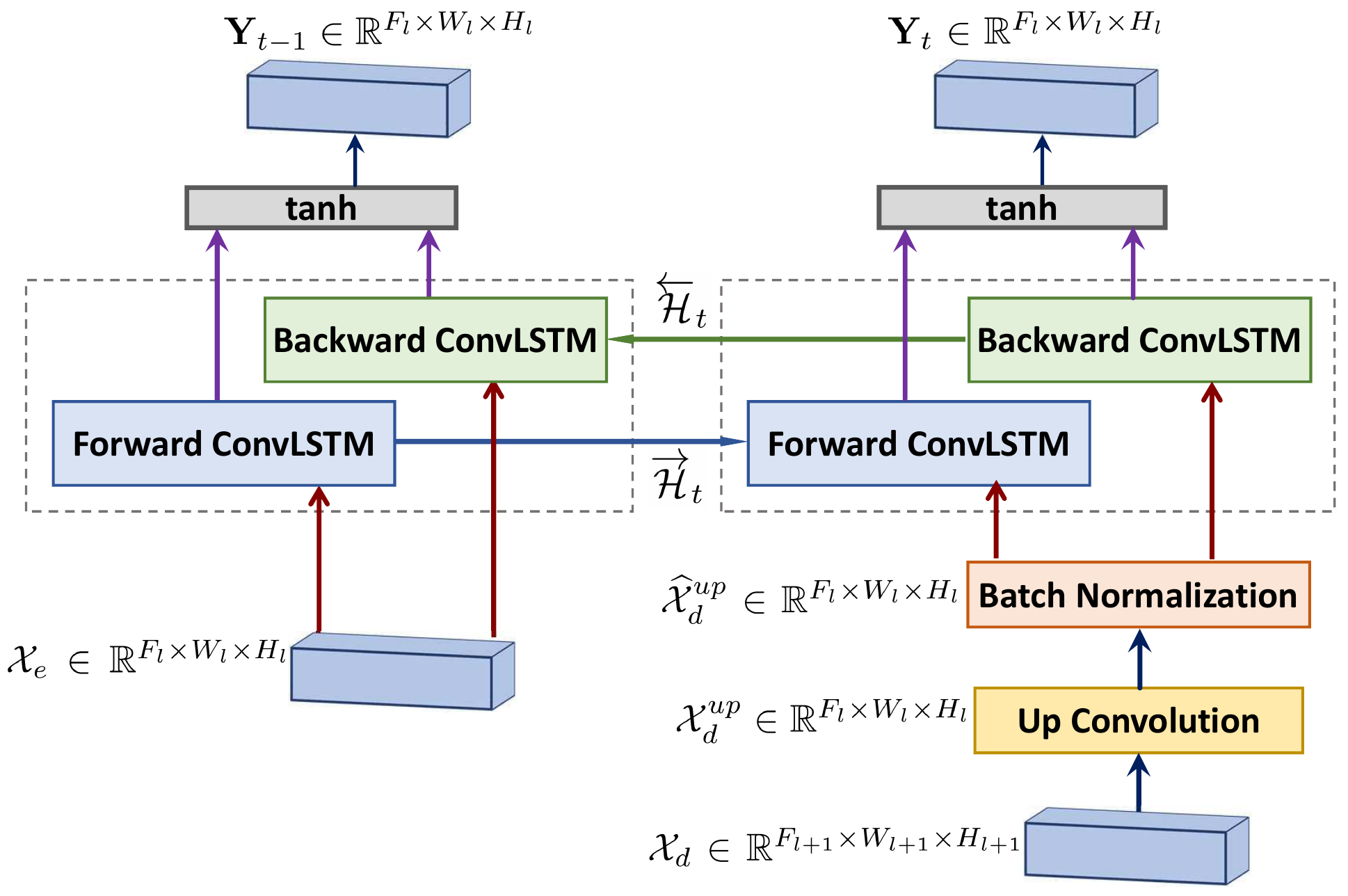}
%\vspace{-4mm}
\caption{Bi-directional ConvLSTM in CUA-Net.} \label{fig:BConvLSTM}
\vspace*{-\baselineskip}
\end{figure}

Each step in the decoding path starts with performing an up-sampling function over the output of the previous layer. In the standard U-Net, the corresponding feature maps in the contracting path are cropped and copied to the decoding path. These feature maps are then concatenated with the output of the up-sampling function. In BCDU-Net, we employ BConvLSTM to process these two kinds of feature maps in a more complex way. 
Let $\mathcal{X}_e \in \mathbb{R}^{F_l\times W_l\times H_l}$ be the set of feature maps copied from the encoding part, and $\mathcal{X}_d\in \mathbb{R}^{F_{l+1}\times W_{l+1}\times H_{l+1}}$ be the the set of feature maps from the previous convolutional layer, where $F_l$ is number of feature maps at layer $l$, and $W_l\times H_l$ is the size of each feature map at layer $l$. It is worth mentioning that $F_{l+1} = 2 *F_{l}$, $W_{l+1} = \frac{1}{2} *W_l$, and $ H_{l+1} = \frac{1}{2} * H_l$. 
Based on Figure \ref{fig:BConvLSTM}, $\mathcal{X}_d$  is first passed to an up-convolutional layer in which an up-sampling function followed by a $2\times2$ convolution are applied, doubling the size of each feature map and halving the number of feature channels, i.e., producing $\mathcal{X}_{d}^{up}\in \mathbb{R}^{F_l\times W_l\times H_l}$. In other words, the expanding path increases the size of the feature maps layer by layer to reach the original size of the input image after the final layer.

%\vspace*{-\baselineskip}
\subsubsection{Batch Normalization:}
After up-sampling, $\mathcal{X}_{d}^{up}$ goes through a BN function and produces $\widehat{\mathcal{X}}_{d}^{up}$. A problem in the intermediate layers in training step is that the distribution of the activations varies. This problem makes the training process very slow since each layer in every training step has to learn to adapt themselves to a new distribution. BN \cite{ioffe2015batch} is utilized to increase the stability of a neural network, which standardizes the inputs to a layer in the network by subtracting the batch mean and dividing by the batch standard deviation. BN affectedly accelerates the speed of training process of a neural network. Moreover, in some cases the performance of the model is improved thanks to the modest regularization effect. More details can be found in \cite{ioffe2015batch}.

\subsubsection{Bi-Directional ConvLSTM:}
The output of the BN step ($\widehat{\mathcal{X}}_{d}^{up} \in \mathbb{R}^{F_l\times W_l\times H_l}$) is now fed to a BConvLSTM layer. 
The main disadvantage of the standard LSTM is that these networks does not take into account the spatial correlation since these models use full connections in input-to-state and state-to-state transitions. To solve this problem, ConvLSTM \cite{xingjian2015} was proposed which exploited convolution operations into input-to-state and state-to-state transitions. It consists of an input gate $i_t$, an output gate $o_t$, a forget gate $f_t$, and a memory cell $\mathcal{C}_t$. Input, output and forget gates act as controlling gates to access, update, and clear memory cell. ConvLSTM can be formulated as follows (for convenience we remove the subscript and superscript from the parameters):

%\vspace*{-\baselineskip}
\begin{equation}
\label{equ:ConvLSTM1}
\centering
\begin{aligned}
    &i_t = \sigma \left( \textbf{W}_{xi} * \mathcal{X}_t + \textbf{W}_{hi} * \mathcal{H}_{t-1} + \textbf{W}_{ci} * \mathcal{C}_{t-1} + b_i \right) \\
    &f_t = \sigma \left( \textbf{W}_{xf}* \mathcal{X}_t + \textbf{W}_{hf}* \mathcal{H}_{t-1} +\textbf{W}_{cf}* \mathcal{C}_{t-1} + b_f \right) \\
    &\mathcal{C}_t = f_t \circ \mathcal{C}_{t-1} + i_t \tanh\left( \textbf{W}_{xc}* \mathcal{X}_t + \textbf{W}_{hc}* \mathcal{H}_{t-1} + b_c \right) \\
    & o_t = \sigma \left( \textbf{W}_{xo}* \mathcal{X}_t + \textbf{W}_{ho}* \mathcal{H}_{t-1} +\textbf{W}_{co}\circ \mathcal{C}_{t} + b_c \right) \\
    &\mathcal{H}_t = o_t \circ \tanh(\mathcal{C}_t),\\
\end{aligned}
\end{equation}
%\vspace*{-\baselineskip}

\noindent where $*$ and $\circ$ denote the convolution and Hadamard functions, respectively. $\mathcal{X}_t$ is the input tensor (in our case $\mathcal{X}_e$ and $\widehat{\mathcal{X}}_{d}^{up}$), $\mathcal{H}_t$ is the hidden sate tensor, $\mathcal{C}_t$ is the memory cell tensor, and, $\textbf{W}_{x*}$ and $\textbf{W}_{h*}$ are 2D Convolution kernels corresponding to the input and hidden state, respectively, and $b_i$, $b_f$, $b_o$, and $b_c$ are the bias terms.

In this network, we employ BConvLSTM \cite{song2018pyramid} to encode $\mathcal{X}_e $ and $\widehat{\mathcal{X}}_{d}^{up}$. BConvLSTM uses two ConvLSTMs to process the input data into two directions of forward and backward paths, and then makes a decision for the current input by dealing with the data dependencies in both directions. In a standard ConvLSTM, only the dependencies of the forward direction are processed. However, all the information in a sequence should be fully considered, therefore, it might be effective to take into account backward dependencies. 
It has been proved that analyzing both forward and backward temporal perspectives enhanced the predictive performance \cite{cui2018deep}.
Each of the forward and backward ConvLSTM can be considered as a standard one. Therefore, we have two sets of parameters for backward and forward states. 
The output of the BConvLSTM is calculated as %$\textbf{Y}_t = \tanh{\left( \textbf{W}_y^{\overrightarrow{\mathcal{H}}} * \overrightarrow{\mathcal{H}}_t +   \textbf{W}_y^{\overleftarrow{\mathcal{H}}} \overleftarrow{\mathcal{H}}_t + b \right)}$ 

%\vspace*{-\baselineskip}
\begin{equation}
\label{equ:ConvLSTM2}
\centering
\begin{aligned}
\textbf{Y}_t = \tanh{\left( \textbf{W}_y^{\overrightarrow{\mathcal{H}}} * \overrightarrow{\mathcal{H}}_t +   \textbf{W}_y^{\overleftarrow{\mathcal{H}}} \overleftarrow{\mathcal{H}}_t + b \right)}\\
\end{aligned}
\end{equation}
\vspace*{-\baselineskip}

\noindent where $\overrightarrow{\mathcal{H}}_t$ and $\overleftarrow{\mathcal{H}}_t$ denote the hidden sate tensors for forward and backward states, respectively, $b$ is the bias term, and $\textbf{Y}_t \in \mathbb{R}^{F_l\times W_l\times H_l}$ indicates the final output considering bidirectional spatio-temporal information. Moreover, $\tanh$ is the hyperbolic tangent which is utilized here to combine the output of both forward and backward states through a non-linear way. We utilize the energy function like the original U-Net to train the network.

% %\vspace*{-\baselineskip}
% \begin{equation}
% \label{equ:ConvLSTM2}
% \centering
% \begin{aligned}
% \textbf{Y}_t = \tanh{\left( \textbf{W}_y^{\overrightarrow{\mathcal{H}}} * \overrightarrow{\mathcal{H}}_t +   \textbf{W}_y^{\overleftarrow{\mathcal{H}}} \overleftarrow{\mathcal{H}}_t + b \right)}\\
% \end{aligned}
% \end{equation}
% \vspace*{-\baselineskip}
%\noindent 

% \subsubsection{Training:}
% Like original U-Net, the energy function utilized in this model is like the original U-Net. A pixel-wise softmax over the last feature map is calculated as $\textbf{P}(y=k) = \nicefrac{\exp{\big(a_k(x) \big)}}{\sum_{k^{\prime}=1}^{K} \exp{\big(a_{k^{\prime}}(x) \big)}}$ such that $k\in\{0,1\}$. where $K$ is the number of classes (2 in our case), $a_k(x)$ is the value of feature channel $k$ at pixel position $x \in \Omega$ with $\Omega \in \mathbb{Z}^2 $. The utilized binary cross-entropy in our model is $E = - \big(y \log{P} + (1-y) \log{(1-P)} \big)$.

% \begin{equation}
% \label{equ:ConvLSTM2}
% \centering
% \begin{aligned}
% &\textbf{P}(y=k) = \frac{\exp{\big(a_k(x) \big)}}{\sum_{k^{\prime}=1}^{K} \exp{\big(a_k^{\prime}(x) \big)}}\\
% &s.t.\ \  k\in\{0,1\}
% \end{aligned}
% \end{equation}

% \begin{equation}
% \label{equ:ConvLSTM3}
% \centering
% \begin{aligned}
% E = - \big(y \log{P} + (1-y) \log{(1-P)} \big)
% \end{aligned}
% \end{equation}

\section{Experimental Results}
We evaluate BCDU-Net on DRIVE, ISIC 2018, and a lung segmentation public benchmark datasets. DRIVE is a dataset for blood vessel segmentation from retina images, ISIC is for skin cancer lesion segmentation, and  the last dataset consists of diagnostic and lung cancer screening thoracic computed tomography (CT) scans with marked-up annotated lesions.
The empirical results show that the proposed method outperforms state-of-the-art alternatives. Keras with TenserFlow backend is utilized for this implementation. %It is worth mentioning that 
The network is trained from scratch for all datasets.
We consider several performance metrics to perform the experimental comparative, including accuracy (AC), sensitivity (SE), specificity (SP), F1-Score, Jaccard similarity (JS), and area under the curve (AUC). We stop the training of the network when the validation loss remains the same in 10 consecutive epochs which is 50, 100, and 25 for DRIVE, ISIC, and Lung datasets, respectively.
%By using variables TP (true positive), FP (false positive), TN (true negative), and FN (false negative), GT (ground truth) and SR (segmentation result) evaluation metrics are calculated as 

% \begin{equation}
% \label{equ:measure}
% \centering
% \begin{aligned}
% &AC = \frac{TP+TN}{TP+FP+TN+FN} \\
% &SE = \frac{TP}{TP+FN} \\
% &SP = \frac{TN}{TN+FP} \\
% &PC = \frac{TP}{TP+FP}\\
% &F1 =2* \frac{PC*SE}{PC+SE}\\
% &JS = \frac{|GR\cap SR|}{|GR\cup SR|}\\
% \end{aligned}
% \end{equation}

% \begin{figure*}[ht]
% 	\centering
% 	\begin{tabular}{ccc}
% 		% Requires \usepackage{graphicx}
% 		\includegraphics[width=0.3\textwidth]{DRIVE_Dataset.png}&
% 		\includegraphics[width=0.3\textwidth]{ISIC_Dataset.png}&
% 		\includegraphics[width=0.3\textwidth]{Lung_Dataset.png}\\
% 		(a) DRIVE, & (b) ISIC, & (c) Lung Segmentation,\\
% 	\end{tabular}
% 	\caption{Some samples of three datasets.}
% 	\label{fig:Datasets}
% \end{figure*}

\subsection{DRIVE Dataset}

DRIVE \cite{staal2004} is a dataset for blood vessel segmentation from retina images. %, shown in Figure \ref{fig:Datasets} (a).
It includes 40 color retina images, from which 20 samples are used for training and the remaining 20 samples for testing. The original size of images is $565\times584$ pixels. It is clear that a dataset with this number of samples is not sufficient for training a deep neural network. Therefore, we use the same strategy as \cite{alom2018} for training our network. The input images are first randomly divided into a number of patches. In total, around $190,000$ patches are produced from $20$ training images, from which $171,000$ patches are used for training, and the remaining $19,000$ patches are used for validation. The size of batches utilized as the input data to the network is $64\times64$. 

% \begin{figure}
% \centering
% \includegraphics[width=0.5\textwidth]{DRIVE_Dataset.png}
% \caption{Some samples of DRIVE dataset.} \label{fig3DRIVE}
% \end{figure}

Some precise and promising segmentation results of the experimental output of the proposed network are shown in Figure \ref{fig:DRIVE_R}. The first column is the original color image, the second one is the ground truth mask and the third column is the output of the proposed BCDU-Net. 
%The training and validation accuracy of the proposed network is shown in Figure \ref{fig:converge} (a) for DRIVE dataset. It can be seen that he network is converged very fast for this dataset.
Table \ref{tab:drive} lists the quantitative results obtained by different methods and the proposed network on DRIVE dataset. %Term "d" in this table means the number of dense block we utilized in the network. 
We evaluate the network with $d=1$ and $d=3$ as the number of dense blocks in the network. With $d=1$ we have one convolutional block without any dense connection in that layer, i.e., like the last encoding layer of the standard U-Net. With $d=3$ we have three convolutional blocks and two dense connections in that layer. 
It is shown that the BCDU-Net (with both $d=1$ and $d=3$) outperforms w.r.t. the state-of-the-art alternatives for most of the evaluation metrics. Moreover, it can be seen that the network with $d=3$ works better than the network without dense block.  

\begin{figure}
\centering
\includegraphics[width=0.35\textwidth]{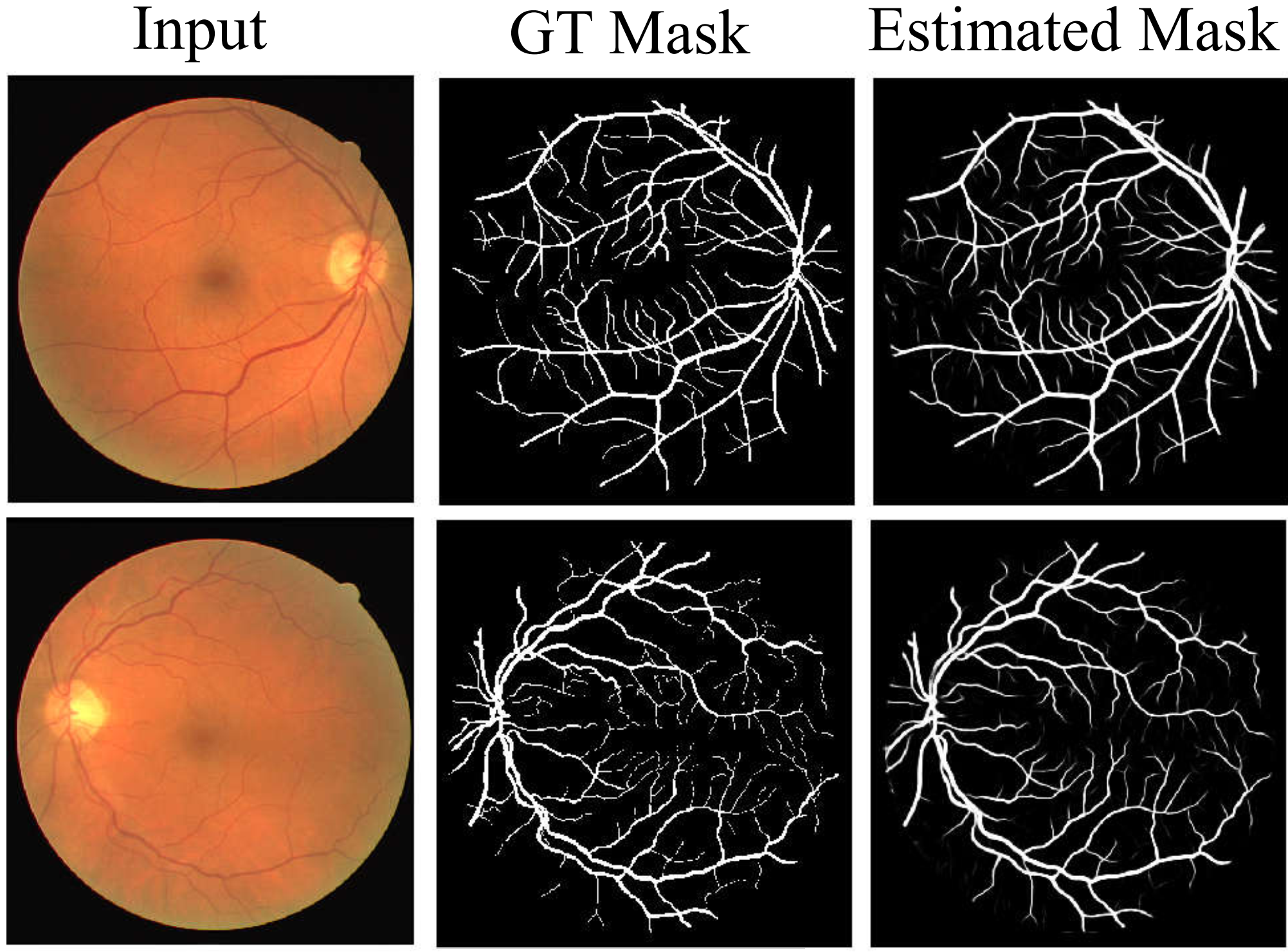}
\caption{Segmentation result of BCDU-Net on DRIVE.} 
\vspace*{-\baselineskip}
\label{fig:DRIVE_R}
\end{figure}

\begin{table*}
\centering
    \vspace*{-\baselineskip}
	\caption{Performance comparison of the proposed network and the state-of-the-art methods on DRIVE dataset.}
	\begin{tabular}{cccccc}
		\hline
		\textbf{Methods} & \textbf{F1-Score}&	\textbf{Sensitivity}&	\textbf{Specificity}&	\textbf{Accuracy}&	\textbf{AUC}\\
		\hline
		%Hybrid Features \cite{cheng2014} &  -	&0.7252&	0.9798&	0.9474&	0.9648\\
		%Three Stage Filtering \cite{roychowdhury2014} & -&	0.7250&	\textbf{0.9830}&	0.9520&	0.9620\\
		COSFIRE filters \cite{azzopardi2015} & -&	0.7655&	0.97048&	0.9442&	0.9614\\
		Cross-Modality \cite{li2015cross} & -&	0.7569&	0.9816&	0.9527&	0.9738\\
		U-net \cite{ronneberger2015} & 0.8142&	0.7537&	0.9820&	0.9531&	0.9755\\
		Deep Model \cite{liskowski2016} & -&	0.7763&	0.9768&	0.9495&	0.9720\\
		%Attention U-net \cite{oktay2018} & 0.8155&	0.7751&	0.9816&	0.9556&	0.9782\\
		RU-net \cite{alom2018} & 0.8149&	0.7726&	0.\textbf{9820}&	0.9553&	0.9779\\
		R2U-Net \cite{alom2018} &0.8171& 0.7792&	0.9813&	0.9556&	0.9782\\
		\hline
		\textbf{BCDU-Net (d=1)}& 0.8222&\textbf{0.8012}&0.9784&	0.9559&  0.9788\\
		\textbf{BCDU-Net (d=3)}& \textbf{0.8224}&0.8007&0.9786&	\textbf{0.9560}&\textbf{0.9789}\\
		\hline
	\end{tabular}
	\label{tab:drive}
\end{table*}
%\vspace*{-\baselineskip}   

\begin{figure*}[ht]
	\centering
	\begin{tabular}{ccc}
		\includegraphics[width=0.31\textwidth,height=30mm]{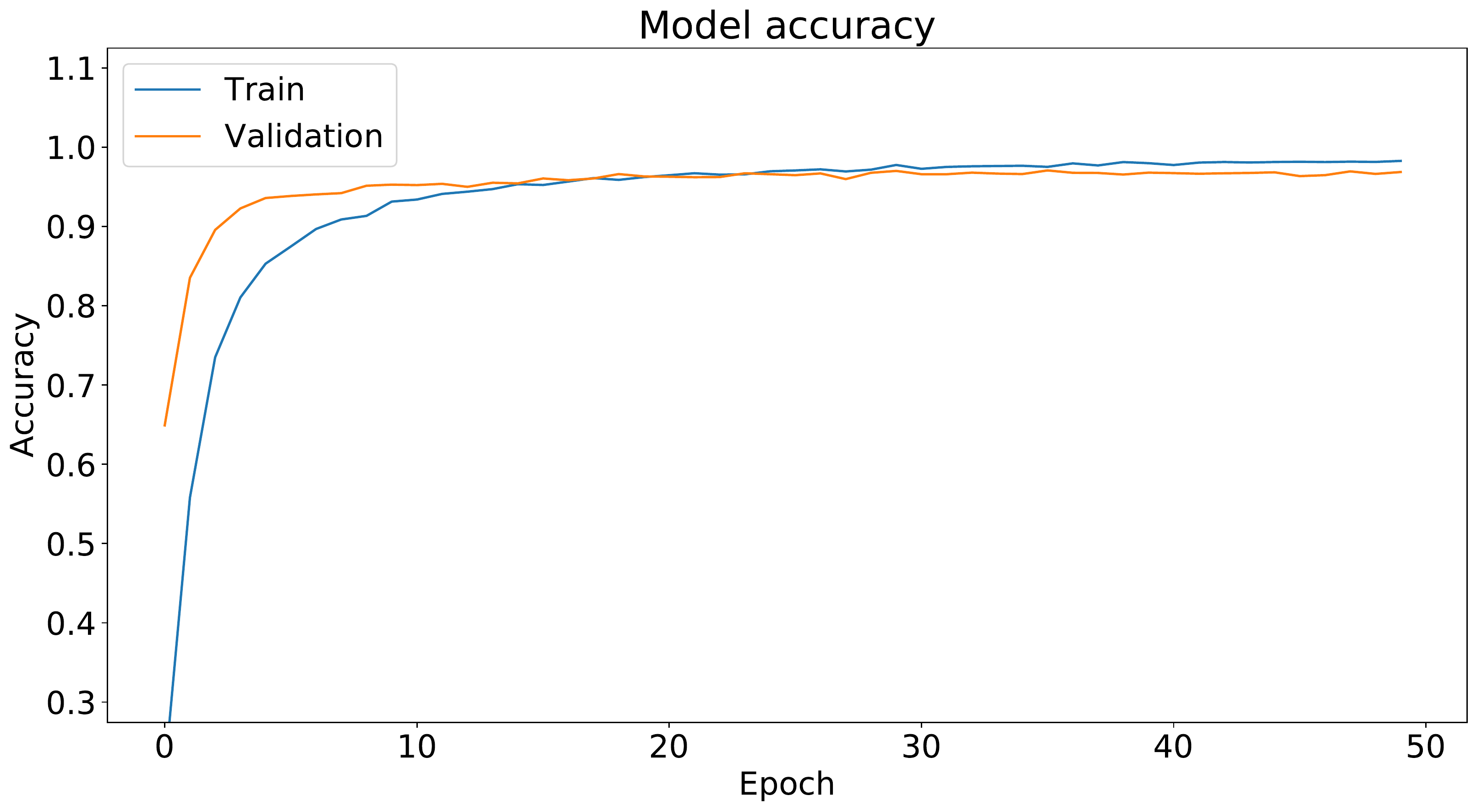}&
		\includegraphics[width=0.31\textwidth,height=30mm]{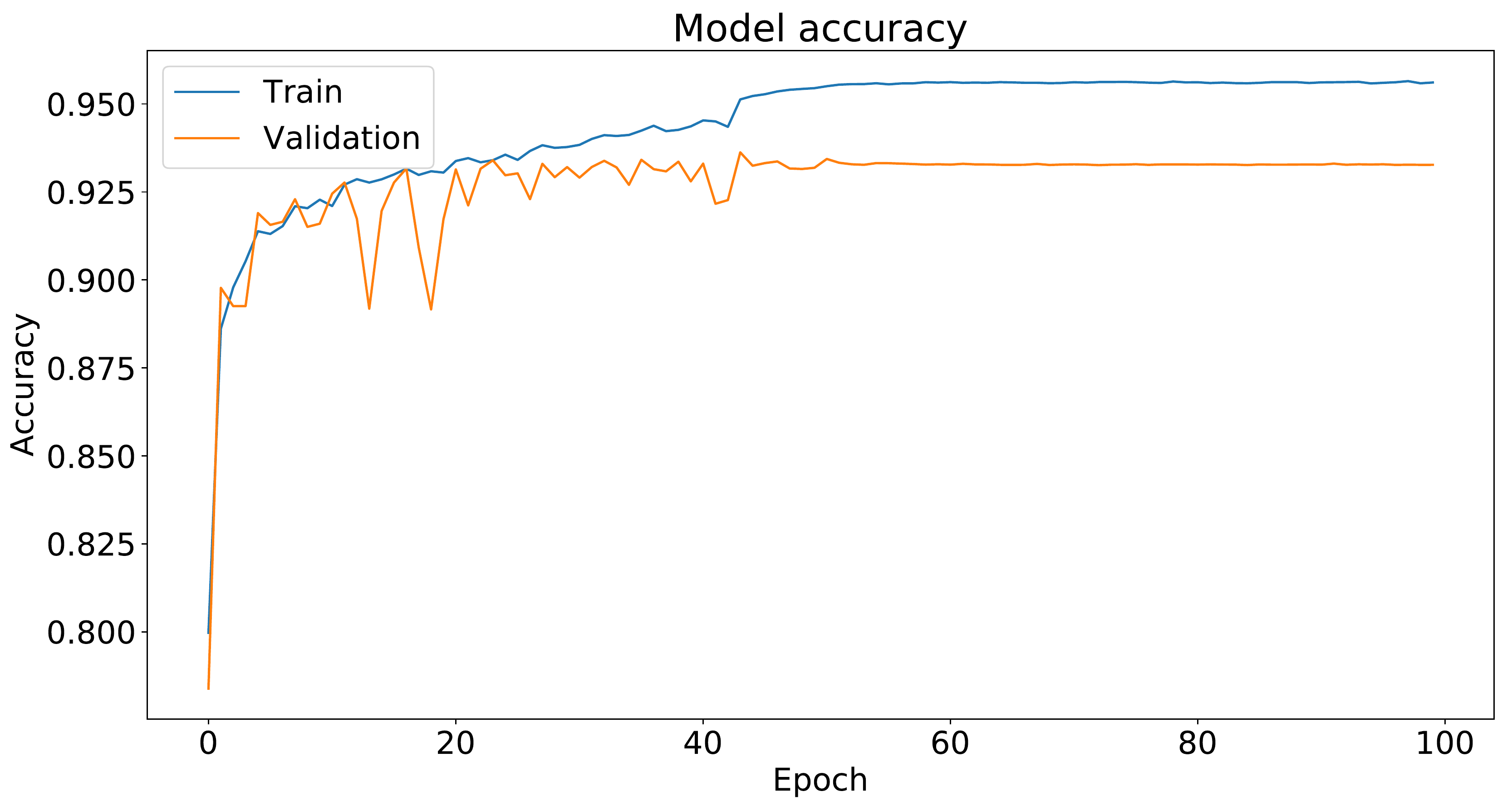}&
		\includegraphics[width=0.31\textwidth,height=30mm]{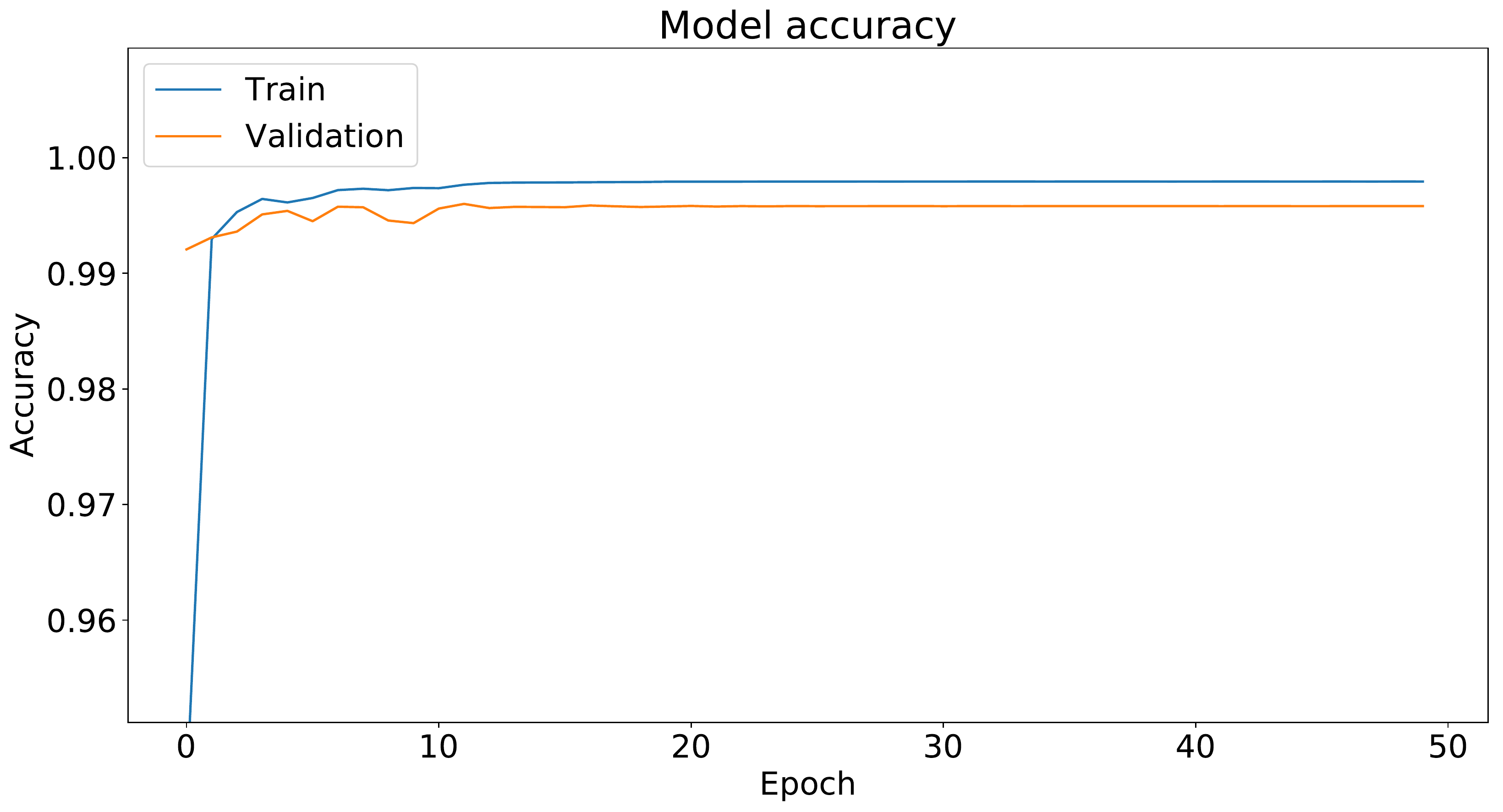}\\
		(a) DRIVE, & (b) ISIC, & (c) Lung Segmentation,\\
	\end{tabular}
	\caption{Training and validation accuracy of BCDU-Net for three datasets.}
	\vspace*{-\baselineskip}
	\label{fig:converge}
\end{figure*}

\begin{figure*}[ht]
	\centering
	\begin{tabular}{ccc}
		\includegraphics[width=0.31\textwidth,height=30mm]{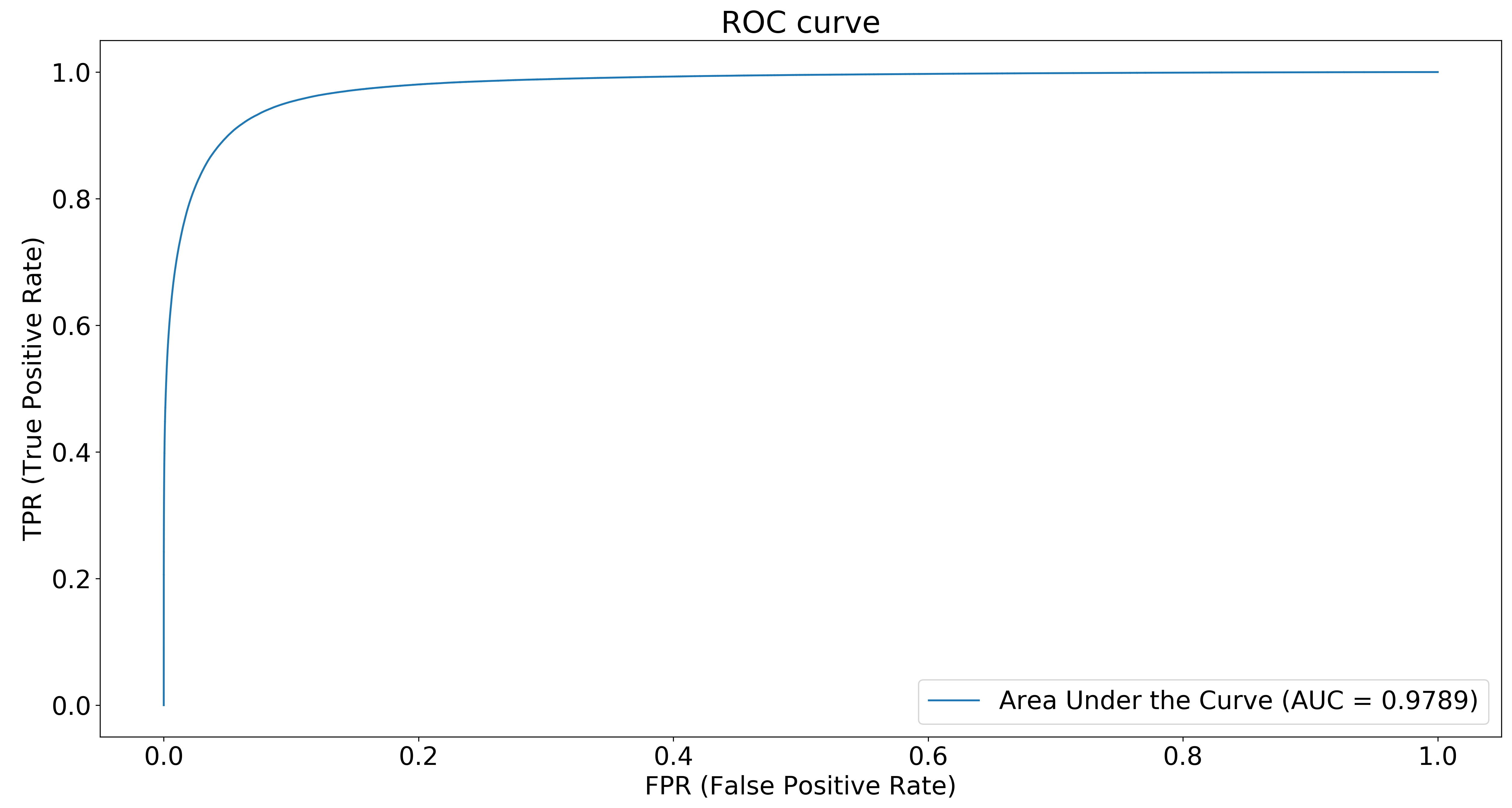}&
		\includegraphics[width=0.31\textwidth,height=30mm]{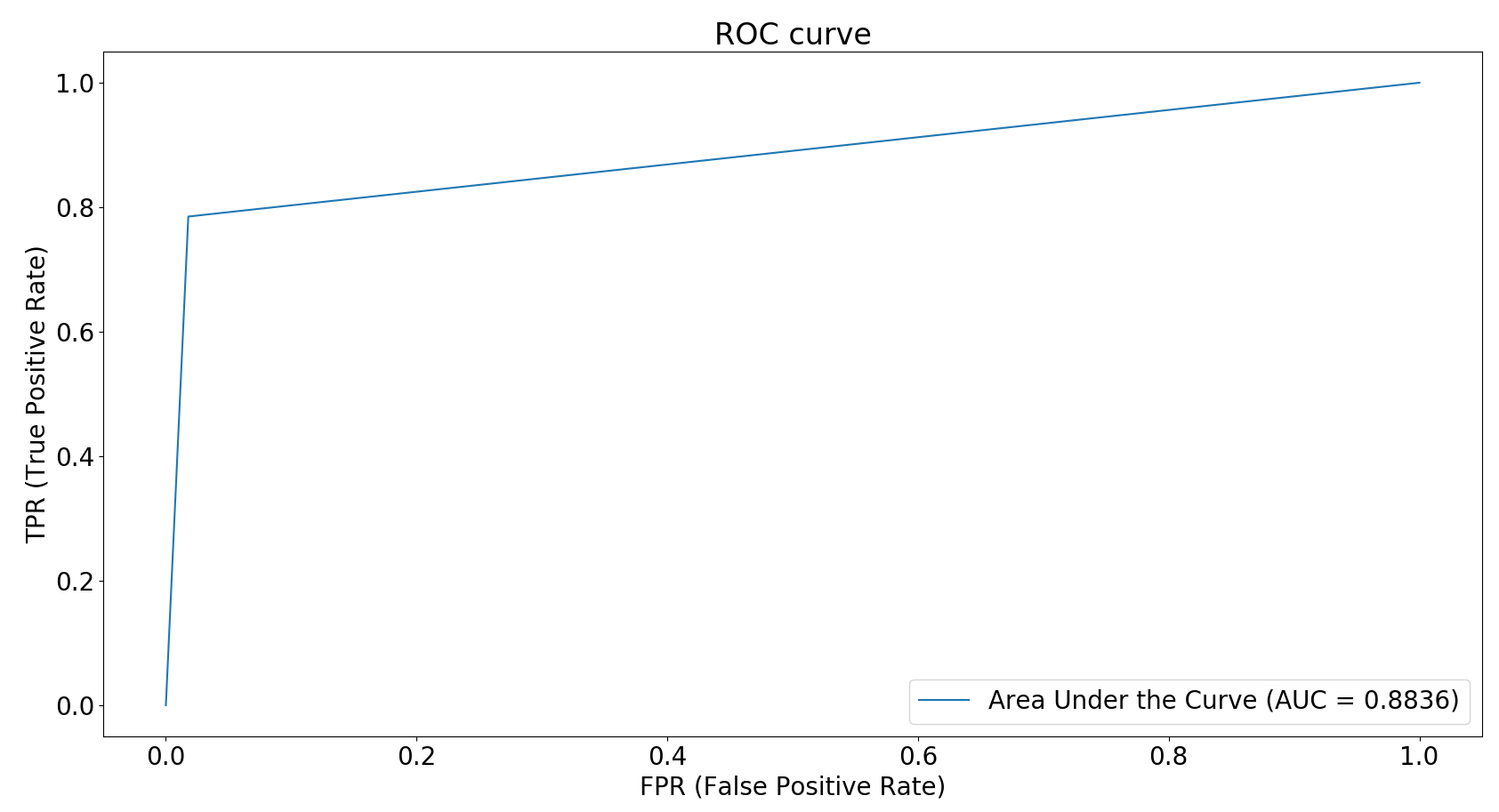}&
		\includegraphics[width=0.31\textwidth,height=30mm]{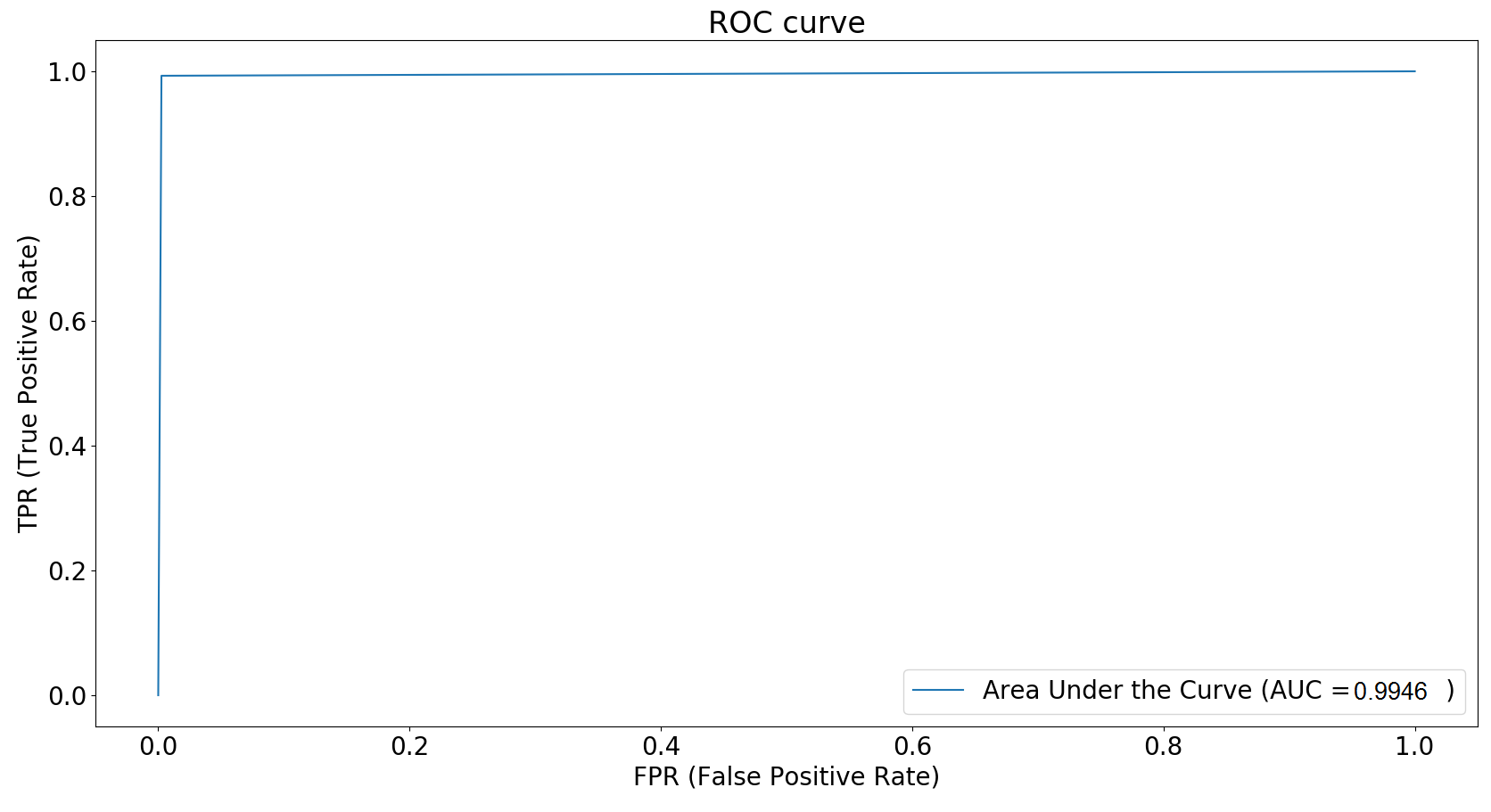}\\
		(a) DRIVE, & (b) ISIC, & (c) Lung Segmentation,\\
	\end{tabular}
	\caption{ROC diagrams of the proposed BCDU-Net for three dataset.}
	\vspace*{-\baselineskip}
	\label{fig:ROCs}
\end{figure*}

To ensure the proper convergence of the proposed network, the training and validation accuracy for DRIVE dataset is shown in Figure \ref{fig:converge} (a). It is shown that the network converges very fast, i.e., after the $30^{th}$ epoch, the network is almost converged. We also can see that in the first 15 epochs the validation accuracy is larger than the training one. This fact is mostly because of the small size of dataset since we use a small set of images as the validation set. Moreover, it might be related to the fact that we evaluate the validation set at the end of epoch. To show the overall performance of the BCDU-Net on DRIVE dataset, ROC curves is shown in Figure \ref{fig:ROCs} (a). ROC is the plot of the true positive rate (TPR) against the false positive rate (FPR). AUC (reported in Table \ref{tab:drive}) is the area under the ROC curve and is a measure of how well the network can segment the input data.

% \begin{table*}
% \centering
% 	\caption{Performance comparison of the proposed network and the state-of-the-art methods on DRIVE dataset.}\label{tab1}
% 	\begin{tabular}{cccccc}
% 		\hline
% 		\textbf{Methods} & \textbf{F1-Score}&	\textbf{Sensivity}&	\textbf{Specificaty}&	\textbf{Accuracy}&	\textbf{AUC}\\
% 		\hline
% 		Hybrid Features (2014) \cite{cheng2014} &  -	&0.7252&	0.9798&	0.9474&	0.9648\\
% 		Three Stage Filtering (2014) \cite{roychowdhury2014} & -&	0.7250&	\textbf{0.9830}&	0.9520&	0.9620\\
% 		COSFIRE filters(2015) \cite{azzopardi2015} & -&	0.7252&	0.9798&	0.9474&	0.9648\\
% 		Cross-Modality (2015) \cite{li2015cross} & -&	0.7569&	0.9816&	0.9527&	0.9738\\
% 		U-net (2015) \cite{ronneberger2015} & 0.8142&	0.7537&	0.9820&	0.9531&	0.9755\\
% 		Deep Model (2016) \cite{liskowski2016} & -&	0.7763&	0.9768&	0.9495&	0.9720\\
% 		Attention U-net (2018) \cite{oktay2018} & 0.8155&	0.7751&	0.9816&	0.9556&	0.9782\\
% 		RU-net (2018) \cite{alom2018} & 0.8149&	0.7726&	0.9820&	0.9553&	0.9779\\
% 		R2U-Net(2018) \cite{alom2018} &0.8155&	0.7751&	0.9816&	0.9556&	0.9782\\
% 		\hline
% 		\textbf{Proposed BCDU-Net}& \textbf{0.8224}&\textbf{0.8007}&0.9786&	\textbf{0.9560}&\textbf{0.9789}\\
% 		\hline
% 	\end{tabular}
% \end{table*}

\begin{table*}
\centering
    %\vspace*{-\baselineskip}
	\caption{Performance comparison of the proposed network and the state-of-the-art methods on ISIC dataset.}
	\begin{tabular}{ccccccc}
		\hline
		\textbf{Methods} & \textbf{F1-Score}&	\textbf{Sensitivity}&	\textbf{Specificity}&	\textbf{Accuracy}&\textbf{PC}&	\textbf{JS}\\
		\hline
		U-net  \cite{ronneberger2015} & 0.647 & 0.708&	0.964&	0.890&	0.779&	0.549\\
		Attention U-net  \cite{oktay2018} & 0.665&	0.717&	0.967&	0.897&	0.787&	0.566\\
		R2U-net  \cite{alom2018} & 0.679&	0.792&	0.928&	0.880&	0.741&	0.581\\
		Attention R2U-Net \cite{alom2018} &0.691&	0.726&	0.971&	0.904&	0.822&	0.592\\
		\hline
		\textbf{BCDU-Net (d=1)}& 0.847& 0.783 &  0.980&	0.936 &0.922 &0.936\\
		\textbf{BCDU-Net (d=3)}& \textbf{0.851}&\textbf{0.785}&\textbf{0.982}&	\textbf{0.937}&\textbf{0.928}&\textbf{0.937}\\
		\hline
	\end{tabular}
	\label{tab:isic}
\end{table*}

% \begin{figure*}[ht]
% 	\centering
% 	\begin{tabular}{ccc}
% 		% Requires \usepackage{graphicx}
% 		\includegraphics[width=0.3\textwidth,height=30mm]{PC_DRIVE.png}&
% 		\includegraphics[width=0.3\textwidth,height=30mm]{PC_ISIC.png}&
% 		\includegraphics[width=0.3\textwidth,height=30mm]{PC_Lung.png}\\
% 		(a) DRIVE, & (b) ISIC, & (c) Lung Segmentation,\\
% 	\end{tabular}
% 	\caption{Precision-Recall diagrams of the proposed BCDU-Net for three dataset.}
% 	\label{fig:PCs}
% \end{figure*}

\subsection{ISIC 2018 Dataset}
The ISIC dataset \cite{codella2018} %, shown in Figure \ref{fig:Datasets} (b), 
was published by the International Skin Imaging Collaboration (ISIC) as a large-scale dataset of dermoscopy images. This dataset is taken from a challenge on lesion segmentation, dermoscopic feature detection, and disease classification. It includes 2594 images where like previous approaches\cite{alom2018}, we used 1815 images for training, 259 for validation and 520 for testing.
The original size of each sample is $700\times900$. We use the same pre-processing as \cite{alom2018} on the input image, and resize images to $256\times256$. The training data consists of the original images and corresponding ground truth annotations (i.e., binary images containing cancer or non-cancer lesions).

\begin{figure}
\centering
\includegraphics[width=0.35\textwidth]{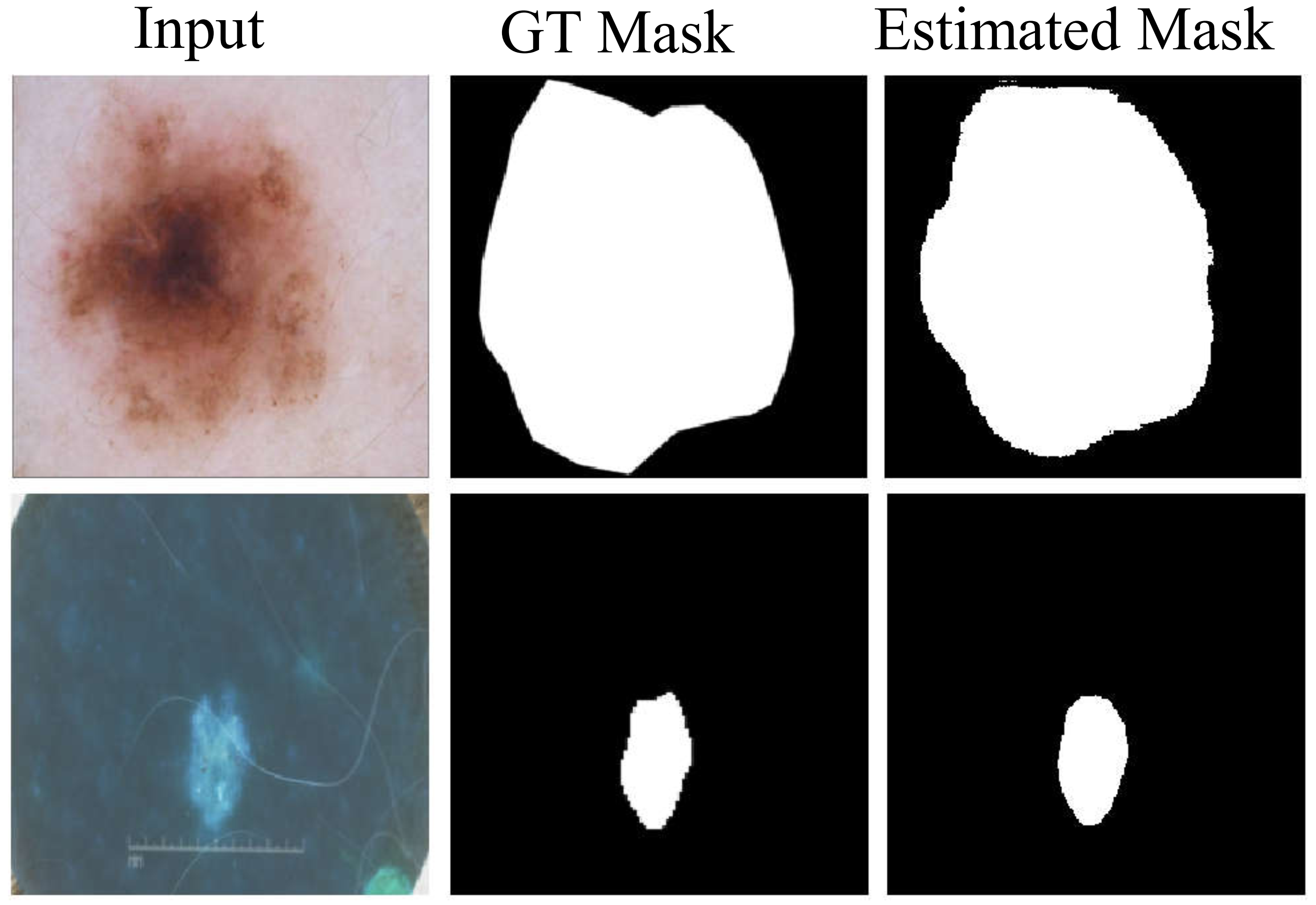}
\caption{Segmentation result of BCDU-Net on ISIC.} 
\vspace*{-\baselineskip}
\label{fig:ISIC_R}
\end{figure}

For qualitative analysis, Figure \ref{fig:ISIC_R} shows some promising example outputs of the proposed BCDU-Net on ISIC dataset. 
Table \ref{tab:isic} lists the quantitative results obtained by different methods and the proposed network on ISIC dataset. A large improvement is achieved by the BCDU-Net (with both $d=1$ and $d=3$ ) w.r.t. state-of-the-art alternatives for all of the evaluation metrics. It is clear that the network with $d=3$ works better than the one with $d=1$.
It is worth mentioning that there was a challenge on ISIC dataset and the best result achieved by the participants was $JS = 0.802$. Compare to this result, there is a good gap between the $JS$ achieved by the BCDU-Net ($0.936$) and the best result of the ISIC challenge.
%Compared to this result, the improvement of BCDU-Net (JS=$0.936$) if more than 0.13 of JS and the best result of the ISIC challenge.
%It is worth mentioning that the best result achieved by the participants in a challenge on ISIC dataset was JS$=0.802$.

The training and validation accuracy of the proposed network for ISIC dataset is shown in Figure \ref{fig:converge} (b). Like DRIVE dataset, the convergence speed of the network for ISIC dataset is fast (after 40 epochs). The validation accuracy over the training process is variable. The reason behind this fact is that the validation set contains some images totally different from the ones in training set, therefore, during the first learning iterations the model has some problems about segmenting those images. To show the overall performance of the BCDU-Net on ISIC dataset, the ROC curves are shown in Figure \ref{fig:ROCs} (b).

% \begin{table*}
% \centering
% 	\caption{Performance comparison of the proposed network and the state-of-the-art methods on ISIC dataset.}\label{tab2}
% 	\begin{tabular}{ccccccc}
% 		\hline
% 		\textbf{Methods} & \textbf{F1-Score}&	\textbf{Sensitivity}&	\textbf{Specificity}&	\textbf{Accuracy}&\textbf{PC}&	\textbf{JS}\\
% 		\hline
% 		U-net (2015) \cite{ronneberger2015} & 0.647 & 0.708&	0.964&	0.890&	0.779&	0.549\\
% 		Attention U-net (2018) \cite{oktay2018} & 0.665&	0.717&	0.967&	0.897&	0.787&	0.566\\
% 		RU-net (2018) \cite{alom2018} & 0.679&	0.792&	0.928&	0.880&	0.741&	0.581\\
% 		R2U-Net(2018) \cite{alom2018} &0.691&	0.726&	0.971&	0.904&	0.822&	0.592\\
% 		\hline
% 		\textbf{Proposed BCDU-Net}& \textbf{0.8506}&\textbf{0.7851}&0.9821&	\textbf{0.9374}&\textbf{0.9280}&\textbf{0.9374}\\
% 		\hline
% 	\end{tabular}
% \end{table*}

% \begin{figure}
% \centering
% \includegraphics[width = 0.5\textwidth]{learning_ISIC.png}
% \caption{Training and validation accuracy of BCDU-Net for ISIC dataset.} 
% \label{figConvergeISIC}
% \end{figure}

% \begin{figure*}[ht]
% 	\centering
% 	\begin{tabular}{cc}
% 		% Requires \usepackage{graphicx}
% 		\includegraphics[width=0.4\textwidth,height=40mm]{Precision_recall_ISIC.png}&
% 		\includegraphics[width=0.4\textwidth,height=40mm]{ROC_ISIC.png}\\
% 		(a) & (b)\\
% 	\end{tabular}
% 	\caption{(a) Precision-Recall and (b) ROC of the BCDU-Net on ISIC dataset .}
% 	\label{fig:driveROC}
% \end{figure*}

\subsection{Lung Segmentation Dataset}
\begin{figure}
\centering
\includegraphics[width=0.35\textwidth]{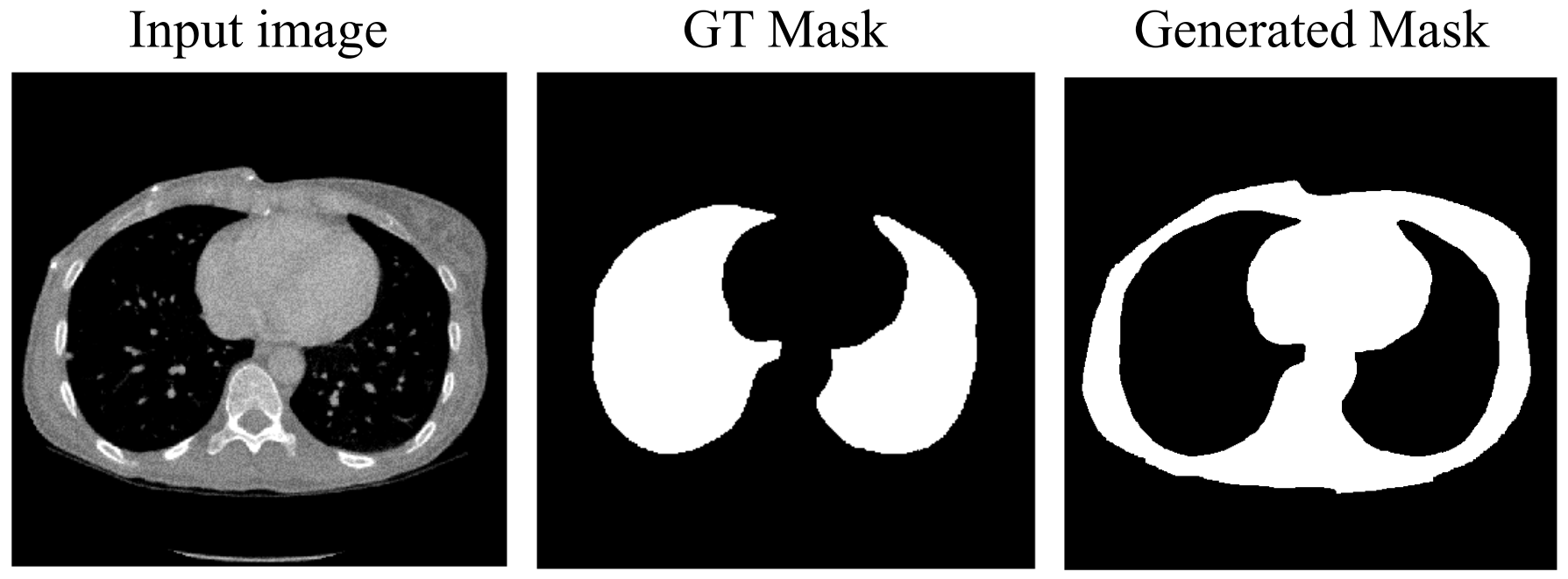}
\caption{A sample of generated mask for Lung dataset.} 
\vspace*{-\baselineskip}
\label{fig:Alg1}
\end{figure}

A lung segmentation dataset is introduced in the Lung Nodule Analysis (LUNA) competition at the Kaggle Data Science Bowl in 2017. This dataset %(shown in Figure \ref{fig:Datasets} (c)) 
consists of 2D and 3D CT images with respective label images for lung segmentation \cite{lungdata}. We use $70\%$ of the data as the train set and the remaining $30\%$ as the test set. The size of each image is $512\times 512$. %, however, we use a resized version of image ($256 \times 256$) as the input data to the proposed network. %For this dataset, we learn the surrounding part of lung tissue, instead of itself. To do that, we apply a pre-processing procedure (shown in Algorithm \ref{Alg:1}) over the input images.
Since the lung region in CT images have almost the same Hausdorff value with non-object of interests such as bone and air, it is worth to learn lung region by learning its surrounding tissues. To do that first we extract the surrounding region by applying algorithm \ref{Alg:1} and then make a new mask for the training sets. We train the model on these new masks and on the testing phase,and estimate the lung region as a region inside the estimated surrounding tissues. A sample is shown in Figure \ref{fig:Alg1} .

%For this dataset, we learn the surrounding part of lung tissue, instead of itself. To do that, we apply a pre-processing procedure (shown in Algorithm \ref{Alg:1}) over the input images. %The whole image consists of lung tissue, bones, and blood vessels. The range of intensity values of pixels is $[-1000,+1000]$. We only need lung tissue of the image, therefore we remove other parts (i.e, bones and blood vessels) by using a threshold ($512$) which is calculates experimentally. The input is then normalized and we convert it to a binary image. Finally, by exploiting morphological functions and the ground truth mask, we create surrounding part of lung tissue.
%\vspace*{-\baselineskip}
\begin{algorithm}
 \caption{Pre-processing over lung dataset.}
 \label{Alg:1}
 \begin{algorithmic}[1]
 \State Input = $X$ and $GT\_Mask$ 
 \State Output = $Surrounding\_Mask$
 \State $X(X>512) = 512$ 
 \NoNumber{$X(X<-512) = -512$}
 \Comment{Remove bones and vessels}{} 
 \State $X = Norm(X)$
 \Comment{Normalize X}{}
 \State $X = image2binary(X)$
 \Comment{Convert to binary}{}
 \State $X = X \cup GT\_Mask$
 \State $X = Morphology (X)$
 \Comment{Remove noise}{}
 \State $Surrounding\_Mask = X-GT\_Mask$
  
 \end{algorithmic}
 \end{algorithm}
 %\vspace*{-\baselineskip}
 
\begin{figure}
\centering
\includegraphics[width=0.35\textwidth]{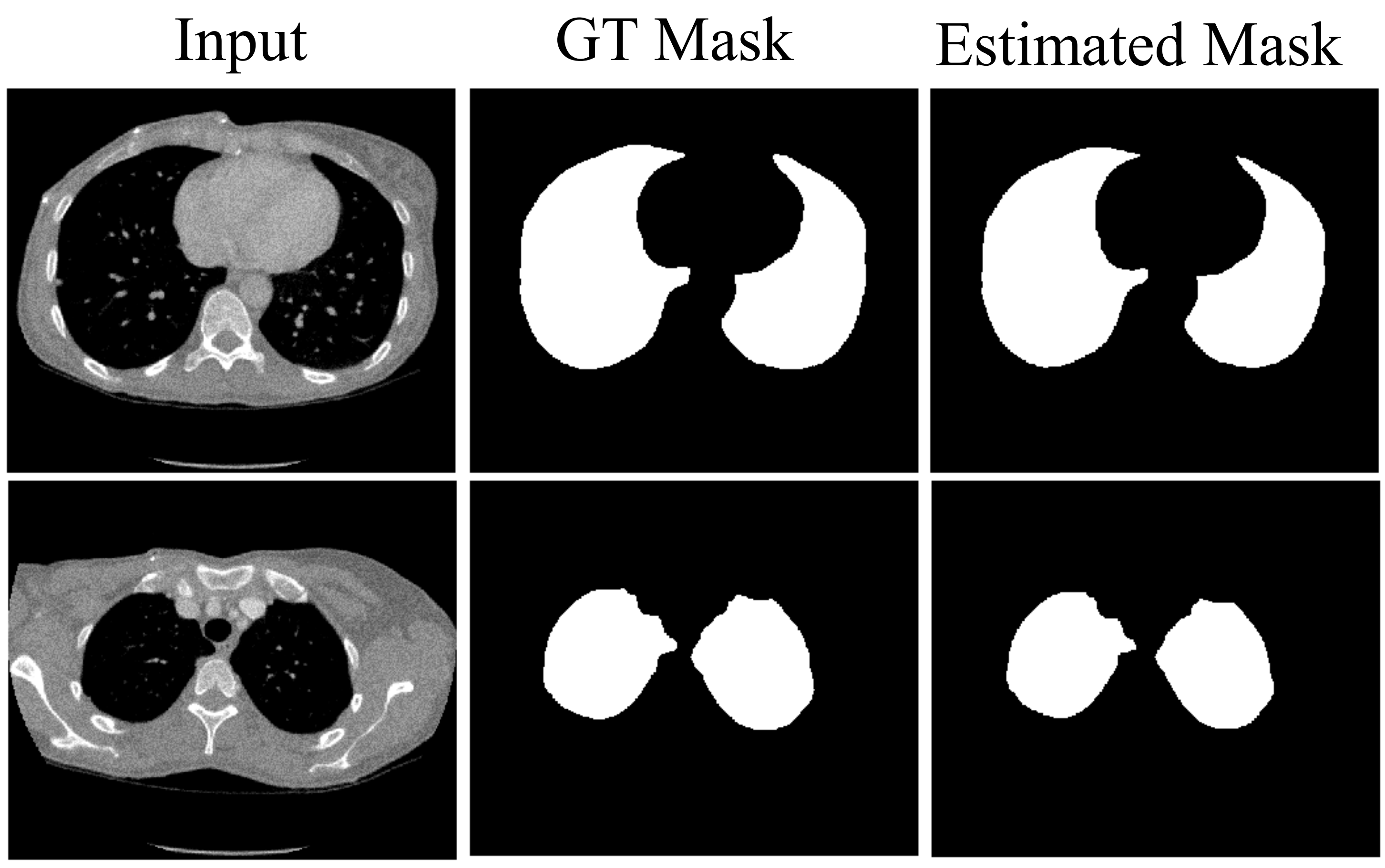}
\caption{Segmentation result of BCDU-Net on Lung dataset.} 
\vspace*{-\baselineskip}
\label{fig:DRIVE_R}
\label{fig:Lung_R}
\end{figure}

Figure \ref{fig:Lung_R} shows some segmentation outputs of the proposed network for lung dataset. The quantitative results of the proposed BCDU-Net is compared with other methods in Table \ref{tab:lung}. It is clear that the BCDU-Net (with both $d=1$ and $d=3$) outperforms the other methods. Moreover, the network with dense connections works better. The training and validation accuracy for this dataset is shown in Figure \ref{fig:converge} (c). To show the overall performance of the network on this dataset, ROC curves is shown in Figure \ref{fig:ROCs} (c).
 \begin{table*}
\centering
    \vspace*{-\baselineskip}
	\caption{Performance comparison of the proposed network and the state-of-the-art methods on Lung dataset.}
	\begin{tabular}{ccccccc}
		\hline
		\textbf{Methods} & \textbf{F1-Score}&	\textbf{Sensitivity}&	\textbf{Specificity}&	\textbf{Accuracy}&\textbf{AUC}&	\textbf{JS}\\
		\hline
		U-net  \cite{ronneberger2015} & 0.9658 & 0.9696&0.9872&	0.9872&	0.9784&	0.9858 \\
		RU-net  \cite{alom2018} &0.9638&	 0.9734&	0.9866&	0.9836&	0.9800&	0.9836 \\
		R2U-Net \cite{alom2018} &0.9832&	\textbf{0.9944}&	0.9832&	0.9918&	0.9889&	0.9918 \\
		\hline
		\textbf{BCDU-Net (d=1)}& 0.9889& 0.9901&0.9979&	0.9967&0.9940&0.9967\\
		\textbf{BCDU-Net (d=3)}& \textbf{0.9904}&0.9910&\textbf{0.9982}&	\textbf{0.9972}&\textbf{0.9946}&\textbf{0.9972}\\
		\hline
	\end{tabular}
	\label{tab:lung}
\end{table*}
 
\subsection{Discussion}
\begin{figure}
%\vspace{-5mm}
	\centering
	\includegraphics[width=0.4\textwidth]{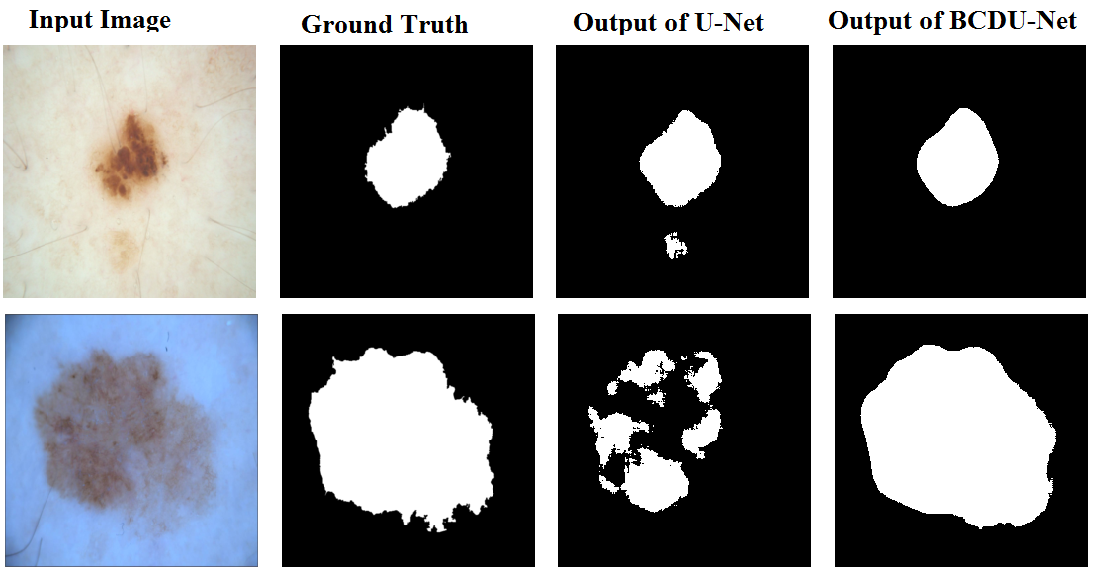}
	\caption{Visual effect of BConvLSTM in BCDU-Net.} 
	 \vspace*{-\baselineskip}
	\label{figConvLSTM}
\end{figure}

The proposed network has some modifications from the original U-Net. We summarized the "Accuracy" and "F1-Score" of the original U-Net and its modifications for three utilized datasets in Table \ref{tab:sum}. We evaluate each modified part of the network to analyze its influence on the result. In Table \ref{tab:sum}, it can be seen that the result of the standard U-Net is improved by inserting BConvLSTM in the skip connections. 
Figure \ref{figConvLSTM} shows the output segmentation mask of the original U-Net and BCDU-Net for two samples of the ISIC dataset. It shows a more precise and fine segmentation output of the proposed network than the original U-Net. 
%In each layer of the expanding path of the original U-Net (and different extensions of this network), 
After the skip connections, there are two kinds of features to combine, i.e., the features from the previous decoding layer and the features from the corresponding encoding layer. 
For convenience, we call them the encoded and decoded features. In the original U-Net, a simple concatenation function is used to combine these two kinds of features.

\begin{table*}
\centering
	\caption{Performance comparison of U-Net and its modifications in our work.}\label{tab2}
	\begin{tabular}{@{\extracolsep{4pt}}ccccccc@{}}
		\hline
		\multirow{2}{*}{\textbf{\textbf{Methods}}}  &\multicolumn{2}{c}{\textbf{DRIVE}} &	\multicolumn{2}{c}{\textbf{ISIC}} &\multicolumn{2}{c}{\textbf{Lung}}\\
		\cline{2-3} \cline{4-5} \cline{6-7}
		& \textbf{F1-Score} & \textbf{AC} & \textbf{F1-Score} & \textbf{AC} & \textbf{F1-Score} & \textbf{AC} \\
		\hline
		U-net  &0.8142 &0.9531 &0.6470 &0.8900 & 0.9658& 0.9828\\
		U-Net + BConvLSTM (d=1) &0.8222 &0.9559 &0.8470 &0.9360 &0.9889 &0.9967 \\
		U-Net + BConvLSTM + Dense Conv (d=3)  &0.8243 &0.9560 &0.8506 & 0.9374& 0.9904& 0.9972\\
		\hline
	\end{tabular}
	\label{tab:sum}
\end{table*}
 %\vspace*{-\baselineskip}

%It can be seen that the result of the standard U-Net is improved by inserting BConvLSTM in the skip connections. Moreover, dense convolution can aff
%\subsubsection{BConvLSTM}

In the proposed network, we used a set of BConvLSTMs to combine encoded and decoded features. The encoded features have higher resolution and therefore contain more local information of the input image, while the decoded features have more semantic information about the input images. The affection of these two features over each other might result in a set of feature maps rich in both local and semantic information. Therefore, instead of a simple concatenation, we utilize BConvLSTM to combine the encoded and decoded features. In BConvLSTM, a set of convolution filters are applied on each kind of features. Therefore each ConvLSTM state, corresponds to one kind of features (e.g. encoded features), ia able to encode relevant information about the other kind of features (e.g. decoded features). %Finally, a non-linear hyperbolic tangent function is employed to combine the output of each ConvLSTM states. 
The convolutional filters along with the hyperbolic tangent functions help the network to learn complex data structures.

\begin{figure}[ht]
	\centering
	%\vspace{-5mm}
	\begin{tabular}{cc}
		\includegraphics[width=0.21\textwidth,height=25mm]{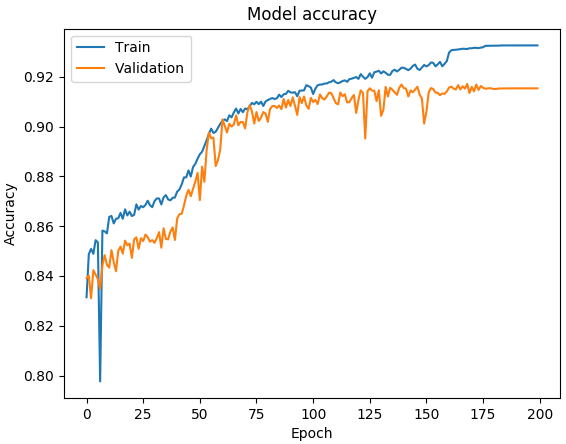}&
		\includegraphics[width=0.21\textwidth,height=25mm]{ISIC.pdf}\\
		(a) & (b)\\
	\end{tabular}
	%\vspace{-5mm}
	\caption{Training and validation accuracy of BCDU-Net (a) without and (b) with BN.}
	\vspace*{-\baselineskip}
	\label{fig:BN}
	%\vspace{-6mm}
\end{figure}

%\subsubsection{Batch Normalization}
We included BN after each up-convolutional layer to speed up the network learning process. To evaluate the effect of this function, we train the network with and without BN. Figure \ref{fig:BN} (a) shows the training and validation accuracy of BCDU-Net for ISIC dataset without BN and Figure \ref{fig:BN} (b) shows the same contend for the network with BN. BCDU-Net converged after 200 epochs without BN while this number is about 30 with BN, i.e., BN yields the network to converge $6.6$ times faster. Moreover, it can be seen that BN has improved the accuracy of the BCDU-Net. The variations among data in the ISIC dataset is larger than the other datasets. BN manages these variations by standardizing data through controlling the mean and variance of distributions of inputs which results in a small regularization and reducing generalization error. Therefore, BN helps the network to improve the performance.

%\subsubsection{Dense Layer}
Table \ref{tab:sum} shows that the network with dense connections improve the accuracy and F1-Score for the three datasets. The key idea of dense convolutions is sharing feature maps between blocks through direct connection between convolutional block. Consequently, each dense block receives all preceding layers as input, and therefore, produces more diversified and richer features. Thus, it helps the network to increase the representational power of deeper models.

We have more feature propagation both in backward and forward paths through dense blocks. The network performs a kind of deep supervision in backward path since dense block receives additional supervision from loss function through shorter connections \cite{huang2017densely}. The error signal is propagated to earlier layers more directly, hence, earlier layers can get direct supervision from the final softmax layer, and moreover, it results in decreasing the vanishing-gradient problem. 
In addition, compared to other deep architectures like residual connections, dense convolutions require fewer parameters while improving the accuracy of the network.

\section{Conclusion}

We proposed BCDU-Net for medical image segmentation. We showed that by including BConvLSTM in the skip connection and also inserting a densely connected convolutional blocks, the network is able to capture more discriminative information which resulted in more precise segmentation results. Moreover, we were able to speed up the network about six times by utilizing BN after the up-convolutional layer. The experimental results on three public benchmark datasets showed high gain in semantic segmentation in relation to state-of-the-art alternatives. %\footnote{Code and trained models will be released in a public github page after publication of the paper.}
%-------------------------------------------------------------------------

\section{Acknowledgment}
This work has been partially supported by the Spanish project TIN2016-74946-P (MINECO/FEDER, UE) and CERCA Programme/Generalitat de Catalunya. We gratefully acknowledge the support of NVIDIA Corporation with the donation of the GPU used for this research. This work is partially supported by ICREA under the ICREA Academia programme.

{\small
\bibliographystyle{ieee}
\bibliography{egbib}

\begin{thebibliography}{10}\itemsep=-1pt

\bibitem{lungdata}
\url{https://www.kaggle.com/kmader/finding-lungs-in-ct-data}.

\bibitem{alom2018}
M.~Z. Alom, M.~Hasan, C.~Yakopcic, T.~M. Taha, and V.~K. Asari.
\newblock Recurrent residual convolutional neural network based on u-net
  (r2u-net) for medical image segmentation.
\newblock {\em arXiv preprint arXiv:1802.06955}, 2018.

\bibitem{azzopardi2015}
G.~Azzopardi, N.~Strisciuglio, M.~Vento, and N.~Petkov.
\newblock Trainable cosfire filters for vessel delineation with application to
  retinal images.
\newblock {\em Medical image analysis}, 19(1):46--57, 2015.

\bibitem{bai2018recurrent}
W.~Bai, H.~Suzuki, C.~Qin, G.~Tarroni, O.~Oktay, P.~M. Matthews, and
  D.~Rueckert.
\newblock Recurrent neural networks for aortic image sequence segmentation with
  sparse annotations.
\newblock In {\em International Conference on Medical Image Computing and
  Computer-Assisted Intervention}, pages 586--594. Springer, 2018.

\bibitem{chen2016voxresnet}
H.~Chen, Q.~Dou, L.~Yu, J.~Qin, and P.-A. Heng.
\newblock Voxresnet: Deep voxelwise residual networks for brain segmentation
  from 3d mr images.
\newblock {\em NeuroImage}, 170:446--455, 2018.

\bibitem{chen2017deeplab}
L.-C. Chen, G.~Papandreou, I.~Kokkinos, K.~Murphy, and A.~L. Yuille.
\newblock Deeplab: Semantic image segmentation with deep convolutional nets,
  atrous convolution, and fully connected crfs.
\newblock {\em IEEE transactions on pattern analysis and machine intelligence},
  40(4):834--848, 2017.

\bibitem{cciccek20163d}
{\"O}.~{\c{C}}i{\c{c}}ek, A.~Abdulkadir, S.~S. Lienkamp, T.~Brox, and
  O.~Ronneberger.
\newblock 3d u-net: learning dense volumetric segmentation from sparse
  annotation.
\newblock In {\em International conference on medical image computing and
  computer-assisted intervention}, pages 424--432. Springer, 2016.

\bibitem{codella2018}
N.~C. Codella, D.~Gutman, M.~E. Celebi, B.~Helba, M.~A. Marchetti, S.~W. Dusza,
  A.~Kalloo, K.~Liopyris, N.~Mishra, H.~Kittler, et~al.
\newblock Skin lesion analysis toward melanoma detection: A challenge at the
  2017 international symposium on biomedical imaging (isbi), hosted by the
  international skin imaging collaboration (isic).
\newblock In {\em ISBI 2018}, pages 168--172. IEEE, 2018.

\bibitem{cui2018deep}
Z.~Cui, R.~Ke, and Y.~Wang.
\newblock Deep bidirectional and unidirectional lstm recurrent neural network
  for network-wide traffic speed prediction.
\newblock {\em arXiv preprint arXiv:1801.02143}, 2018.

\bibitem{cui2016}
Z.~Cui, J.~Yang, and Y.~Qiao.
\newblock Brain mri segmentation with patch-based cnn approach.
\newblock In {\em 2016 35th Chinese Control Conference (CCC)}, pages
  7026--7031. IEEE, 2016.

\bibitem{drozdzal2016}
M.~Drozdzal, E.~Vorontsov, G.~Chartrand, S.~Kadoury, and C.~Pal.
\newblock The importance of skip connections in biomedical image segmentation.
\newblock In {\em Deep Learning and Data Labeling for Medical Applications},
  pages 179--187. Springer, 2016.

\bibitem{huang2017densely}
G.~Huang, Z.~Liu, L.~Van Der~Maaten, and K.~Q. Weinberger.
\newblock Densely connected convolutional networks.
\newblock In {\em Proceedings of the IEEE conference on computer vision and
  pattern recognition}, pages 4700--4708, 2017.

\bibitem{ioffe2015batch}
S.~Ioffe and C.~Szegedy.
\newblock Batch normalization: Accelerating deep network training by reducing
  internal covariate shift.
\newblock pages 448--456, 2015.

\bibitem{kleesiek2016}
J.~Kleesiek, G.~Urban, A.~Hubert, D.~Schwarz, K.~Maier-Hein, M.~Bendszus, and
  A.~Biller.
\newblock Deep mri brain extraction: a 3d convolutional neural network for
  skull stripping.
\newblock {\em NeuroImage}, 129:460--469, 2016.

\bibitem{li2015cross}
Q.~Li, B.~Feng, L.~Xie, P.~Liang, H.~Zhang, and T.~Wang.
\newblock A cross-modality learning approach for vessel segmentation in retinal
  images.
\newblock {\em IEEE transactions on medical imaging}, 35(1):109--118, 2015.

\bibitem{liskowski2016}
P.~Liskowski and K.~Krawiec.
\newblock Segmenting retinal blood vessels with deep neural networks.
\newblock {\em IEEE transactions on medical imaging}, 35(11):2369--2380, 2016.

\bibitem{long2015fully}
J.~Long, E.~Shelhamer, and T.~Darrell.
\newblock Fully convolutional networks for semantic segmentation.
\newblock In {\em Proceedings of the IEEE conference on computer vision and
  pattern recognition}, pages 3431--3440, 2015.

\bibitem{milletari2016}
F.~Milletari, N.~Navab, and S.-A. Ahmadi.
\newblock V-net: Fully convolutional neural networks for volumetric medical
  image segmentation.
\newblock In {\em 2016 Fourth International Conference on 3D Vision (3DV)},
  pages 565--571. IEEE, 2016.

\bibitem{oktay2018}
O.~Oktay, J.~Schlemper, L.~L. Folgoc, M.~Lee, M.~Heinrich, K.~Misawa, K.~Mori,
  S.~McDonagh, N.~Y. Hammerla, B.~Kainz, et~al.
\newblock Attention u-net: Learning where to look for the pancreas.
\newblock {\em arXiv preprint arXiv:1804.03999}, 2018.

\bibitem{pinheiro2014}
P.~O. Pinheiro and R.~Collobert.
\newblock Recurrent convolutional neural networks for scene labeling.
\newblock Technical report, 2014.

\bibitem{ronneberger2015}
O.~Ronneberger, P.~Fischer, and T.~Brox.
\newblock U-net: Convolutional networks for biomedical image segmentation.
\newblock In {\em International Conference on Medical image computing and
  computer-assisted intervention}, pages 234--241. Springer, 2015.

\bibitem{roth2015deeporgan}
H.~R. Roth, L.~Lu, A.~Farag, H.-C. Shin, J.~Liu, E.~B. Turkbey, and R.~M.
  Summers.
\newblock Deeporgan: Multi-level deep convolutional networks for automated
  pancreas segmentation.
\newblock In {\em International conference on medical image computing and
  computer-assisted intervention}, pages 556--564. Springer, 2015.

\bibitem{song2018pyramid}
H.~Song, W.~Wang, S.~Zhao, J.~Shen, and K.-M. Lam.
\newblock Pyramid dilated deeper convlstm for video salient object detection.
\newblock In {\em Proceedings of the European Conference on Computer Vision
  (ECCV)}, pages 715--731, 2018.

\bibitem{staal2004}
J.~Staal, M.~D. Abr{\`a}moff, M.~Niemeijer, M.~A. Viergever, and
  B.~Van~Ginneken.
\newblock Ridge-based vessel segmentation in color images of the retina.
\newblock {\em IEEE transactions on medical imaging}, 23(4):501--509, 2004.

\bibitem{visin2016reseg}
F.~Visin, M.~Ciccone, A.~Romero, K.~Kastner, K.~Cho, Y.~Bengio, M.~Matteucci,
  and A.~Courville.
\newblock Reseg: A recurrent neural network-based model for semantic
  segmentation.
\newblock In {\em Proceedings of the IEEE Conference on Computer Vision and
  Pattern Recognition Workshops}, pages 41--48, 2016.

\bibitem{xingjian2015}
S.~Xingjian, Z.~Chen, H.~Wang, D.-Y. Yeung, W.-K. Wong, and W.-c. Woo.
\newblock Convolutional lstm network: A machine learning approach for
  precipitation nowcasting.
\newblock In {\em Advances in neural information processing systems}, pages
  802--810, 2015.

\bibitem{zhou2016three}
X.~Zhou, T.~Ito, R.~Takayama, S.~Wang, T.~Hara, and H.~Fujita.
\newblock Three-dimensional ct image segmentation by combining 2d fully
  convolutional network with 3d majority voting.
\newblock In {\em Deep Learning and Data Labeling for Medical Applications},
  pages 111--120. Springer, 2016.

\bibitem{zhou2017fixed}
Y.~Zhou, L.~Xie, W.~Shen, Y.~Wang, E.~K. Fishman, and A.~L. Yuille.
\newblock A fixed-point model for pancreas segmentation in abdominal ct scans.
\newblock In {\em International Conference on Medical Image Computing and
  Computer-Assisted Intervention}, pages 693--701. Springer, 2017.

\end{thebibliography}
}

\end{document}